\documentclass[
    ,final            % use final for the camera ready runs
  ]
  {aipproc}

\layoutstyle{6x9}

\newcommand{\spom} {\mbox{$\scriptstyle \mathrm{I}\! \mathrm{P}$}}
\newcommand{\diff} {\mbox{$\scriptstyle \mathrm{diff}$}}

\newcommand{\qsq}      {\ensuremath{Q^2}}
\newcommand{\ttra}     {\ensuremath{|t|}}

\newcommand{\gev}      {\ensuremath{\rm GeV}}
\newcommand{\gevsq}    {\ensuremath{\rm GeV^2}}
\newcommand{\gevsqinv} {\ensuremath{\rm GeV^{-2}}} 

\newcommand{\piplus}   {\ensuremath{\pi^+}}

\newcommand{\rhoz}     {\ensuremath{\rho^{0}}}

\newcommand{\mvm}      {\ensuremath{m_{V}}}

\newcommand{\jpsi}     {\ensuremath{J/\psi}}

\newcommand{\qqbar}    {\ensuremath{q\bar{q}}}

\newcommand{\Pom}      {\ensuremath{I\!\!P}}              
\newcommand{\pom}      {\ensuremath{I\!\!P}}

\newcommand{\apom}     {\ensuremath{\alpha_{\Pom}}}

\newcommand{\aprim} {\ensuremath{\alpha {'}}}

\newcommand{\rzqzz}    {\ensuremath{r_{00}^{04}}}

\newcommand{\rzqumu}   {\ensuremath{r_{1-1}^{04}}}

\begin{document}

\title{Summary of the ``Diffraction $\&$ Vector Mesons'' working group 
at DIS05}


\author{X. Janssen}{
  address={DESY, 22607 Hamburg, Germany}
}

\author{M. Ruspa}{
  address={University of Eastern Piedmont, 28100 Novara, Italy}
}

\author{V. A. Khoze}{
  address={IPPP, University of Durham, DH1 3LE,UK}
}

\begin{abstract}

We survey the contributions presented in the working group ``Diffraction 
$\&$ Vector Mesons'' 
at the XIII International Workshop on Deep Inelastic Scattering 
(http://www.hep.wisc.edu/dis05).

\end{abstract}

\maketitle

%%%%%%%%%%%%%%%%%%%%%%%%%%%%%%%%%%%%%%%%%%%%
%% MAINMATTER
%%%%%%%%%%%%%%%%%%%%%%%%%%%%%%%%%%%%%%%%%%%%

\section{Introduction}

In diffractive interactions in hadron-hadron or photon-hadron 
collisions at least one of the beam particles emerges intact from 
the collision, having lost only a small fraction of its initial energy, 
and carrying a small transverse momentum. 
Therefore no color is exchanged in the $t$-channel. 
The signature for such processes is the presence 
of a gap in rapidity between the two hadronic final states. 
At high energy this is described by the exchange of an object with the 
quantum numbers of the vacuum, referred to 
as the Pomeron in the framework of Regge phenomenology~\cite{regge}. 
Note that at low energies similar reactions can also proceed when quantum 
numbers are exchanged through subleading Regge trajectories (Reggeons); 
however, these contributions are 
exponentially suppressed as a function of the gap size and are negligible 
at small values of the longitudinal momentum loss. The understanding and description of diffractive processes is one of the aims of QCD. 

\vspace{0.2cm}

Diffractive events are being extensively studied at HERA, TEVATRON, RHIC, 
JLAB and CERN and there is a growing community planning to continue 
these studies at the LHC. Updates on the available experimental data 
and on their theoretical interpretation were given at this workshop; 
many discussions also took place on the future plans.  
In the present summary we focus on the path from HERA to the LHC through 
the TEVATRON. 

\section{From HERA to hadron colliders}

\subsection{Selection of diffractive processes}

\begin{figure}
  \includegraphics[height=.2\textheight]{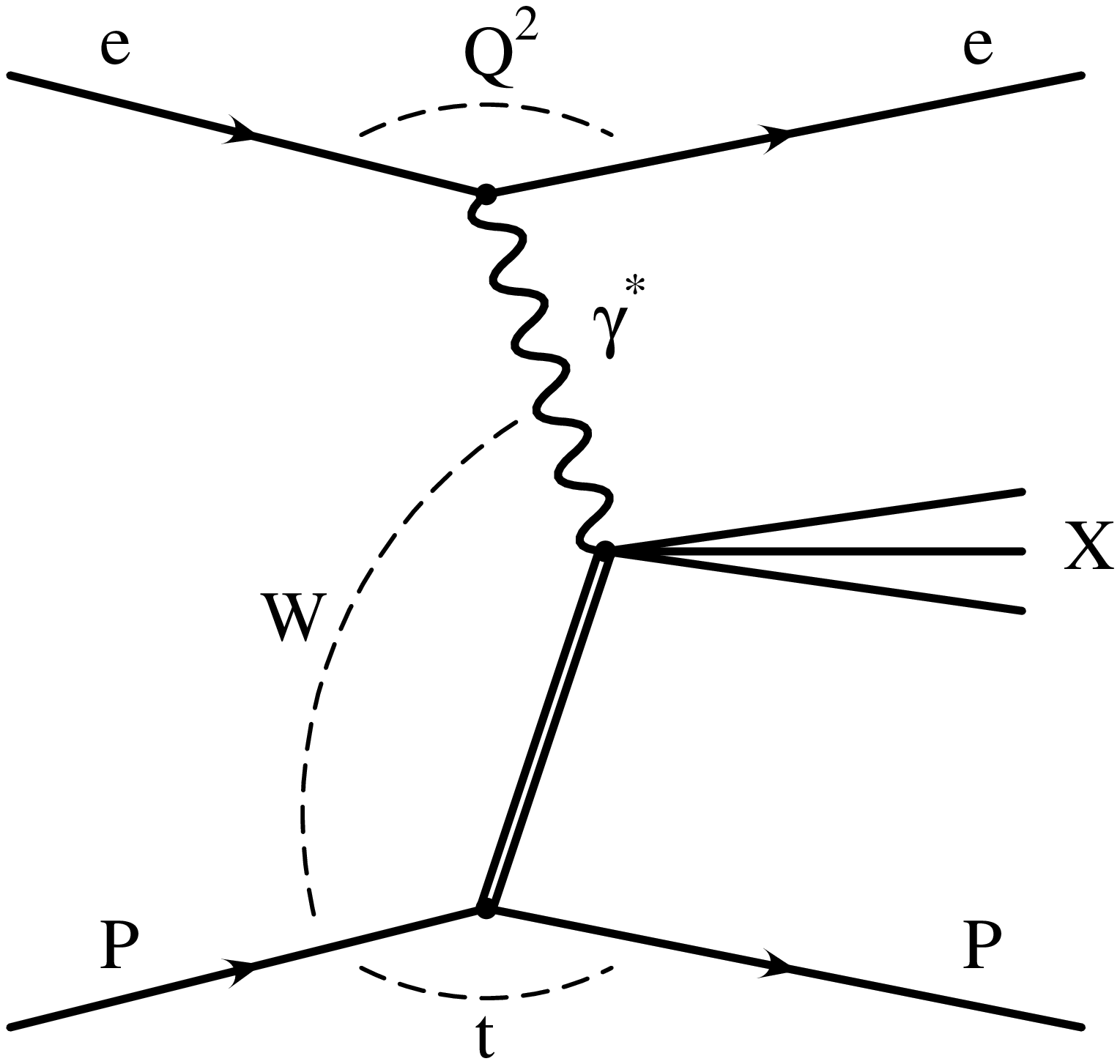}
  \hspace{2cm}
  \includegraphics[height=.2\textheight]{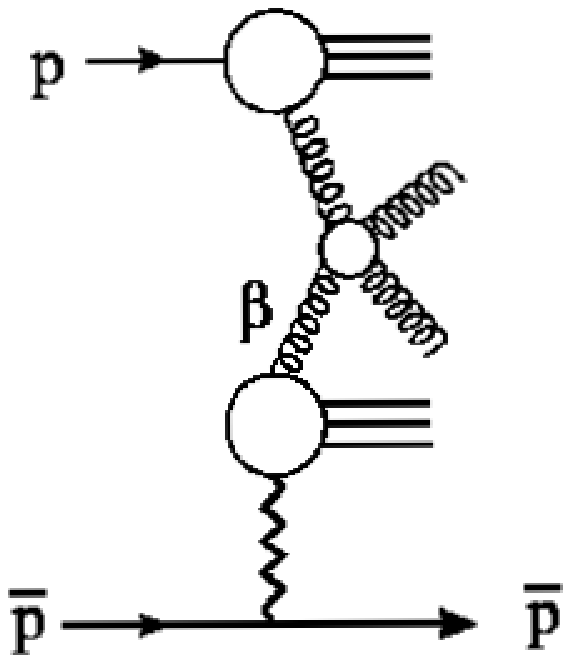}
  \put(-220,33){(a)}
  \put(-25,33){(b)}
  \caption{(a) Diffraction in $ep$ interactions.
           (b) Diffraction in $p\bar{p}$ interactions.}
  \label{fig:f1}
\end{figure}

Let us first look at the diffractive reaction $ep \rightarrow eXp$ at HERA, 
depicted in Fig.~\ref{fig:f1}a: a photon of virtuality $Q^2$ diffractively 
dissociates interacting with the proton at a 
center of mass energy $W$ and squared four momentum transfer $t$ and produces  
the hadronic system $X$ with mass $M_X$ in the final state. 
The fraction of the proton momentum carried by the exchanged object is 
denoted by $x_{\spom}$, while the fraction of the momentum of the exchanged 
object carried by the struck quark is denoted by $\beta$ (note that sometimes 
$z$ is used instead of $\beta$). The virtual photon emitted 
from the lepton beam provides a point-like probe to study the structure of 
the diffractive exchange, similarly to ordinary DIS probing proton structure. 
The fact that a large fraction ($\sim 10\%$) of deep inelastic 
(DIS) events at HERA is diffractive has thus opened the possibility of 
investigating the partonic nature of the Pomeron and has established 
a theoretical link between Regge theory and QCD.  

At the TEVATRON inclusive diffraction is mainly studied via the reaction 
$p\bar{p} \rightarrow \bar{p}X$, sketched in Fig.~\ref{fig:f1}b; 
in the TEVATRON jargon $x_{\spom}$ is usually indicated as $\xi$.

\vspace{0.2cm}

At HERA three methods are used to select diffractive events~\cite{zeuslps}. 
The first is based on the measurement of the scattered proton with a 
spectrometer installed very close to the beam in a region with acceptance for 
protons which have lost only a small fraction of their initial longitudinal 
energy.
A second method requires the presence of a large rapidity gap (LRG) 
in the forward region. A third method is based on the different shape 
of the $M_X$ 
distribution between diffractive and non diffractive events. At the 
TEVATRON diffractive interactions are selected by tagging 
events by either a rapidity gap or a leading antiproton~\cite{tev_sel}.

The proton tagging method 
has the advantage of excluding the proton dissociation processes 
$ep \rightarrow eXN$, where the proton also diffractively dissociates into a 
state $N$ of mass $M_N$ that escapes undetected into the beam pipe. 
%Also higher values of $M_X$ and lower values of $x_{\spom}$ are achieved.
In order to ensure that the scattered proton resulted from a diffractive 
process one requires $x_{\spom} < 0.01$. This cut removes contributions coming 
from Reggeon exchanges~\cite{trajectories}. 

The large rapidity gap method selects events which include some proton 
dissociation processes and some Reggeon contributions. The latter can be 
removed by the same $x_{\spom}$ cut as above. If the mass $M_N$ of the 
dissociative system is large enough to be measured in the forward 
detector the proton dissociation background can be removed, whereas the 
contribution of low mass proton dissociation can be estimated with a 
Monte Carlo simulation ($10\%$ of background with $M_N < 1.6$~GeV is 
quoted from the H1 analysis~\cite{h1lrg}).

\begin{figure}
  \includegraphics[height=.55\textheight]{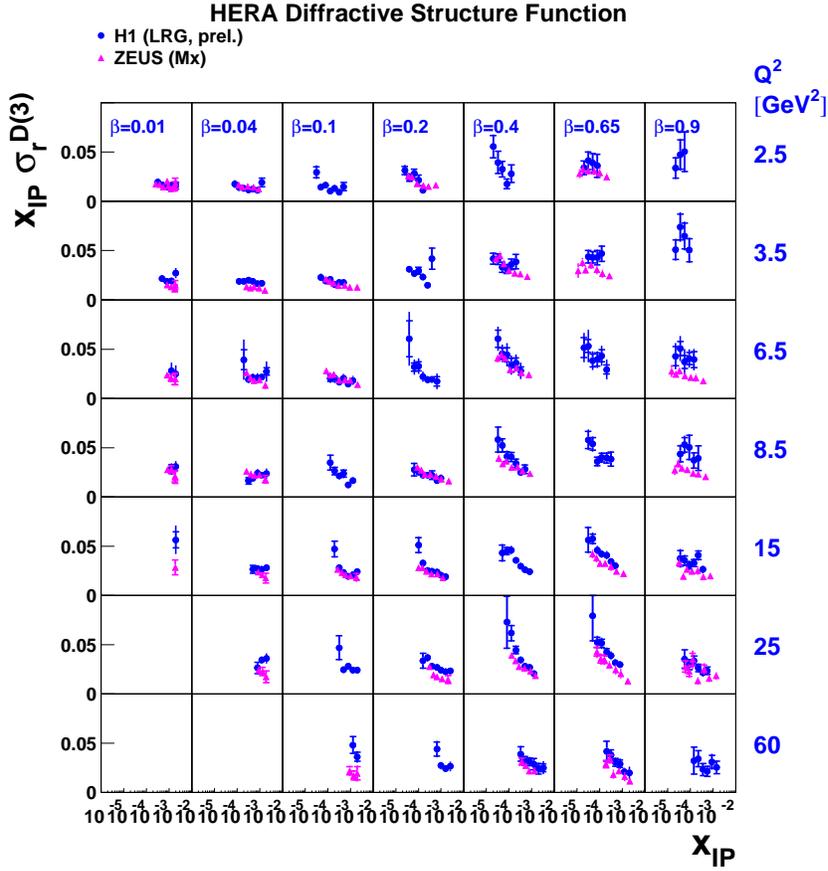}
  \caption{ZEUS $M_X$ and H1 LRG measurements of the diffractive structure 
function.}
  \label{fig:f2}
\end{figure}

In the $M_X$ method the statistical subtraction of the non diffractive 
background eliminates also the the Reggeon 
contribution, but the selected sample is left with an important 
contamination from proton dissociative events with masses 
$M_N < 2.3$~GeV~\cite{zeusmx}. 
By comparing the measured cross sections with those coming from the 
leading proton analysis one can 
estimate the amount of this background (around $30\%$~\cite{lps95}) and 
determine a correction factor. 

\subsection{HERA diffractive structure function and PDFs}

H1 and ZEUS have presented recent precise measurements of the diffractive 
structure function obtained with all 
three HERA methods and covering a wide kinematic range 
(proton tagging method:~\cite{zeuslps,h1fps}, 
 LRG method:~\cite{h1lrg}, 
 $M_x$ method:~\cite{zeusmx}). 
In Fig.~\ref{fig:f2} 
the diffractive structure function is presented as a function of $x_{\spom}$ 
for fixed values of $Q^2$ and $\beta$. 
The data points come from two samples analysed 
by H1 with the LRG method and by ZEUS with the $M_X$ method, respectively. 
The ZEUS $M_X$ data have been 
scaled to $M_Y < 1.6$~GeV, the region of dissociative masses included in the 
H1 data.  
%CORRECTION TO THE H1 DATA? Note that the ZEUS LPS data cover a large range in 
%$x_{\spom}$.  
There is a reasonable agreement between the two data sets, but 
at a closer inspection it turns out that the $Q^2$ dependences are different, 
namely the positive scaling violations in the ZEUS data are smaller than 
in the H1 data. This discrepancy has been investigated very recently by 
a combined set of 
next-to-leading-order (NLO) QCD fits of the diffractive structure function, 
attempted by two different groups (P. Newman et al.~\cite{laycock} and 
A. Levy et al.~\cite{levy} - see also the upcoming proceedings of the 
HERA-LHC workshop, http://www.desy.de/$\sim$heralhc).  

Such fits are based on the validity of a collinear factorization theorem in 
diffractive processes~\cite{fact}, which allows $F_2^D$ to be written 
as a convolution of the usual partonic cross sections as in
 DIS with Diffractive Parton Distribution Functions (DPDFs). The DPDFs, 
parametrised at a starting scale, are evolved according to the  
DGLAP equations~\cite{dglap} and fitted to the data.   
%to use the DGLAP~\cite{dglap}  
%equations to evolve the parton distribution functions (PDFs) parametrised at 
%a starting scale, and to get them once fitted to the data. 
In the ideal case we would evolve in $Q^2$ for fixed $t$ and $x_{\spom}$, 
or at least  
for fixed $x_{\spom}$ if $t$ is integrated over, but this is not allowed by 
the rather limited statistics of the present data. 
%An alternative approach is 
%the Regge factorization hypothesis which factorises
%the $x_{\spom}$ dependence is in a Pomeron flux term. Whether Regge 
%factorization 
An alternative approach is the assumption, known as ``Regge factorization" 
hypothesis, that $F_2^D$ can be expressed as the product of a flux, 
depending only on $x_{\spom}$ and $t$, and the structure function of a 
particle-like object. Whether the data support this assumption or not is a 
controversial problem. It translates into determining 
whether or not the intercept $\apom(0)$ of the Pomeron 
trajectory  $\apom(t) = \apom(0) + \aprim t$ depends on $Q^2$. 

\begin{figure}
  \includegraphics[height=.33\textheight]{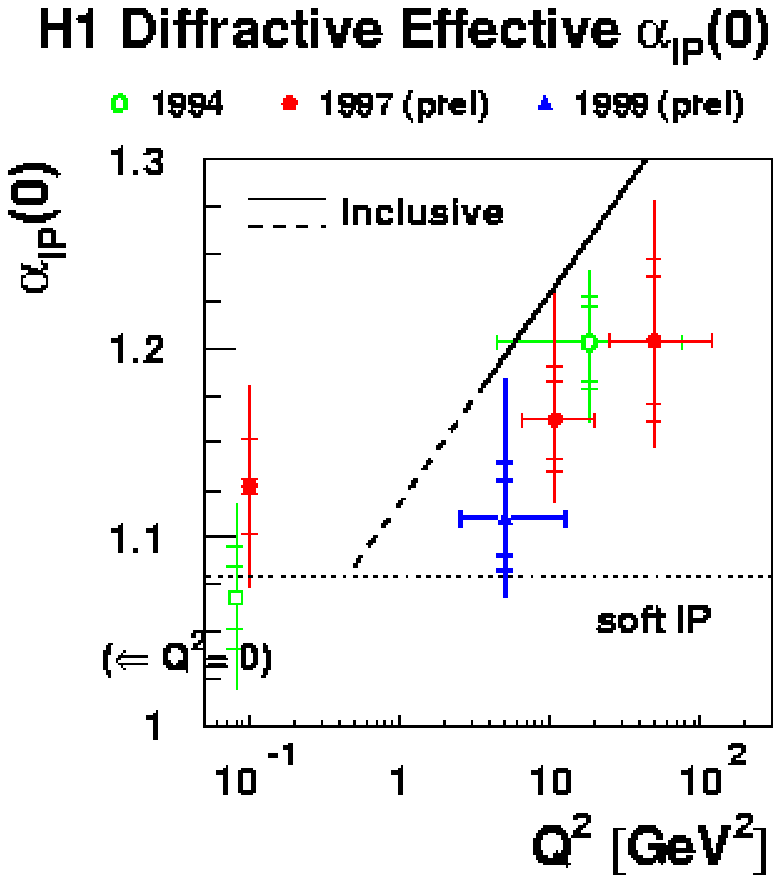}
  \hspace{-1cm}
  \includegraphics[height=.29\textheight]{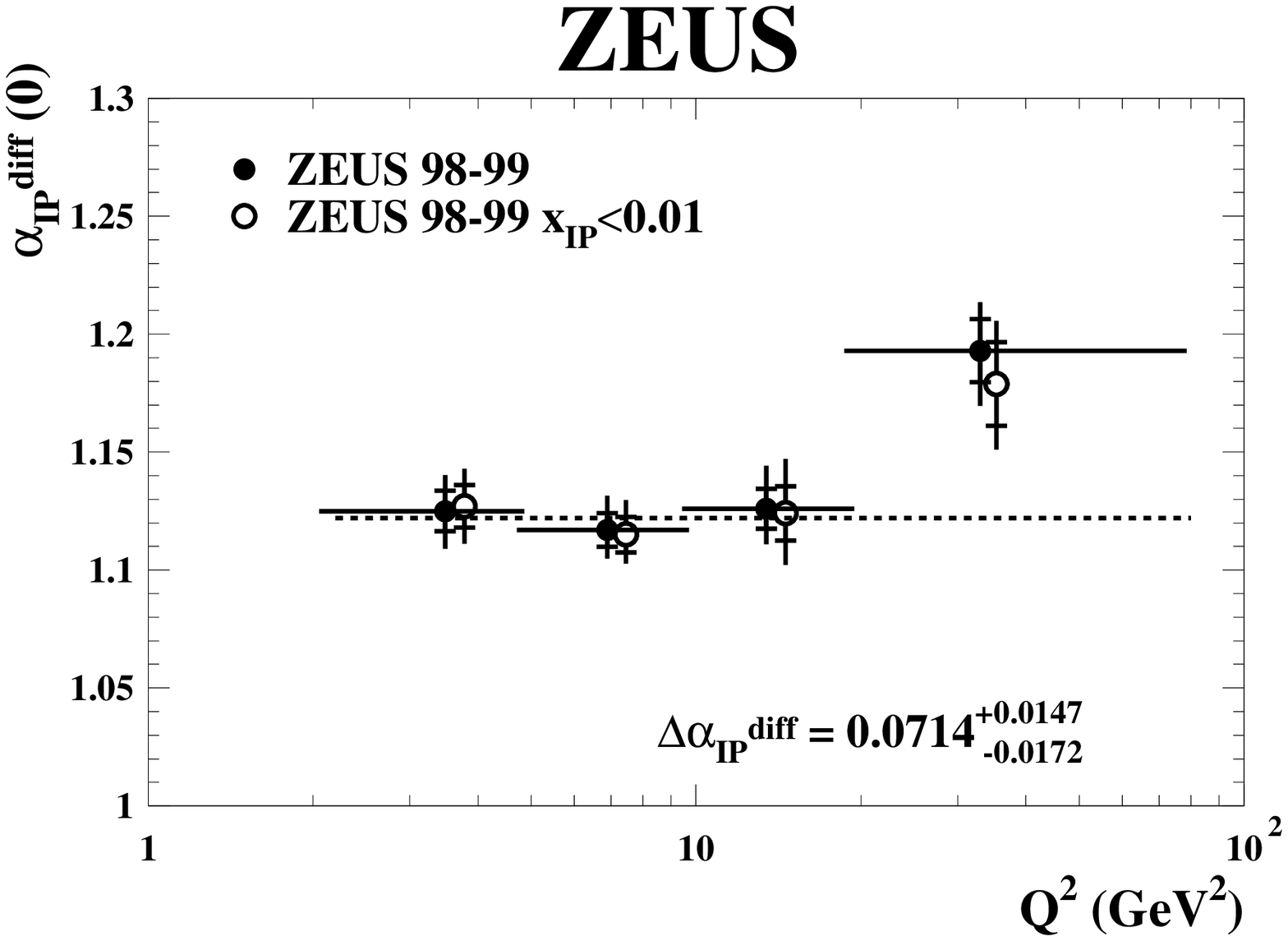}
  \put(-340,5){(a)}
  \put(-200,5){(b)}
  \caption{$Q^2$ dependence of the Pomeron intercept $\apom(0)$, 
           measured by H1 (a) and by ZEUS (b).}
  \label{fig:f3}
\end{figure}

Fig.~\ref{fig:f3}a shows $\apom(0)$ as a function of 
$Q^2$, as measured by H1 : 
there is a suggestion of a dependence of $\apom(0)$ on $Q^2$, 
though firm conclusions are not possible with the present uncertainties. 
In Fig.~\ref{fig:f3}b, where the ZEUS measurement is presented, the 
Pomeron intercept rises by 
$\Delta\alpha_{\diff}=0.0741\pm 0.0140$(stat.)$^{+0.0047}_{-0.0100}$(syst.)
between $Q^2$ of 7.8~GeV$^2$ and 27~GeV$^2$, with a significance of 4.2 
standard deviations. This scenario suggests a possible violation of Regge 
factorization and a clear need for more precise data. 
Nevertheless it has been shown~\cite{levy} that when restricting the 
analysed range to $x_{\spom}~<~0.01$ 
Regge factorization is a sufficiently good 
approximation, and this is the compromise at the basis of the 
NLO DGLAP fits discussed in the following. 

\begin{figure}
  \includegraphics[height=.35\textheight]{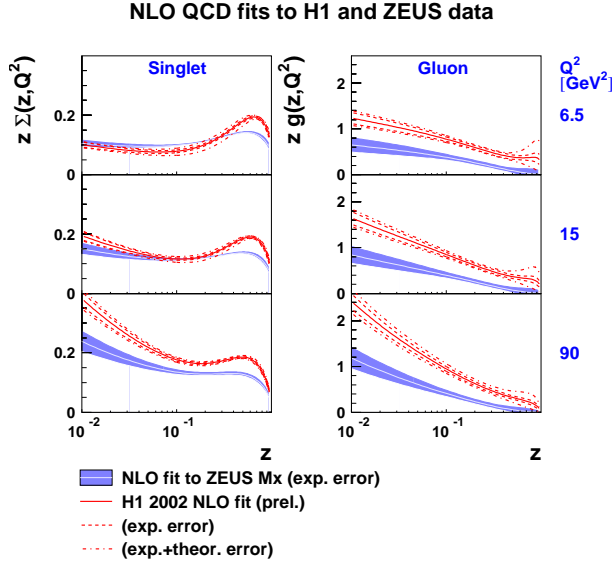}
  \caption{The diffractive parton densities resulting from a NLO QCD 
           fit by P. Newman et al.~\cite{laycock} 
	   to the ZEUS $M_X$ data (solid line) and to the H1 LRG data 
	   (shaded line).}
  \label{fig:f4}
\end{figure}

\begin{figure}
  \includegraphics[height=.21\textheight]{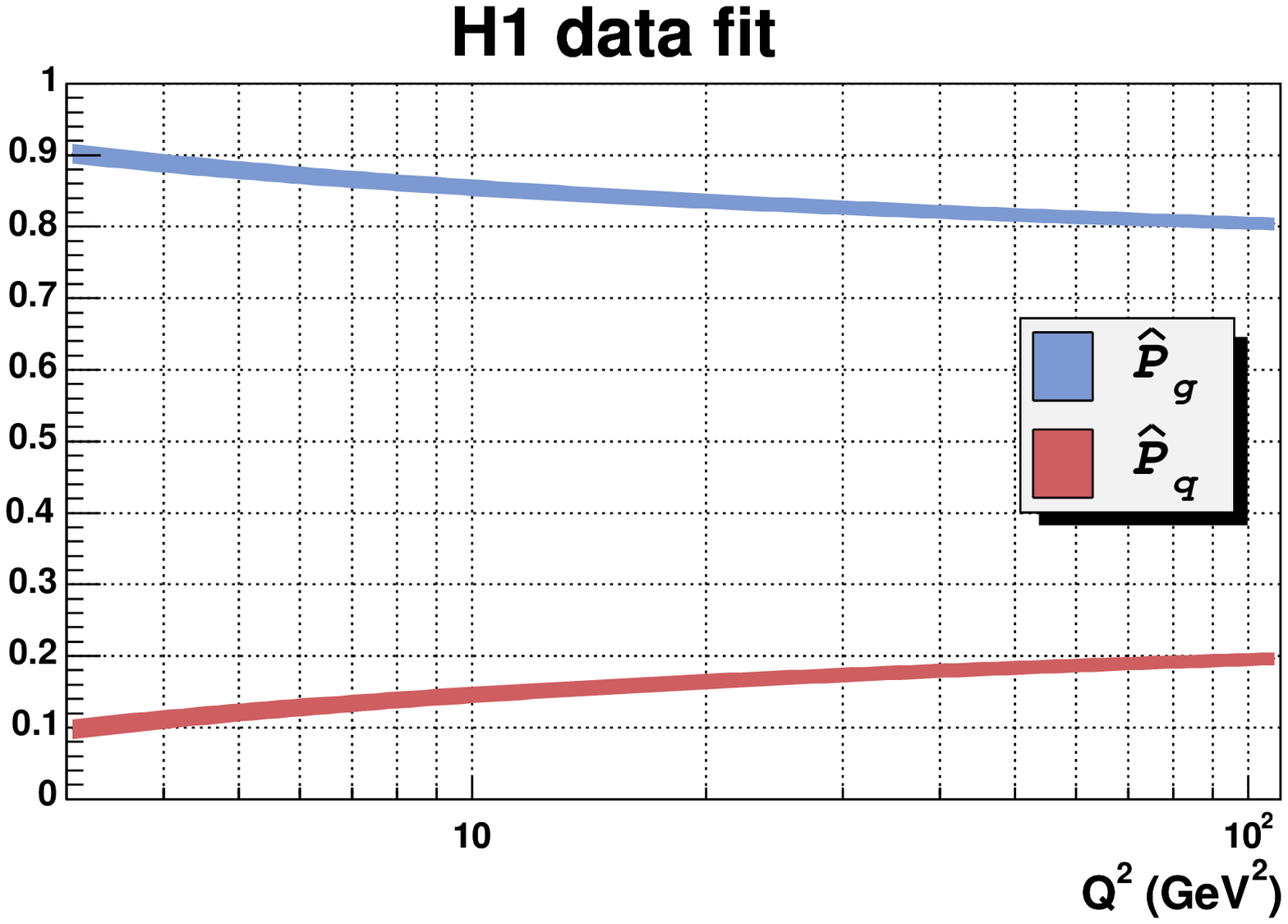}
  \hspace{-1.2cm}
  \includegraphics[height=.21\textheight]{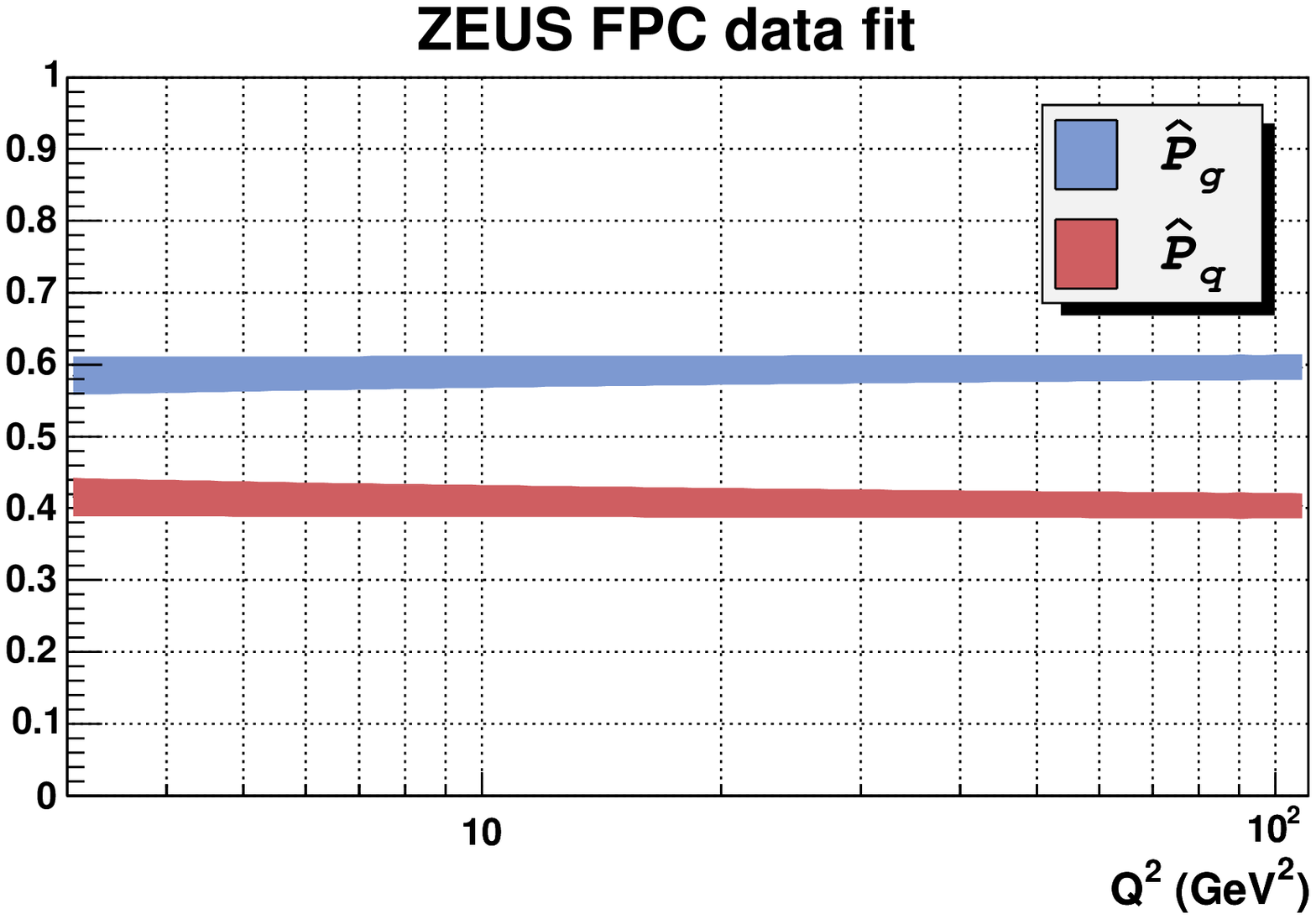}
  \put(-300,2){(a)}
  \put(-120,2){(b)}
  \caption{The parton momentum fraction as a function of $Q^2$ 
           from a NLO QCD fit by A. Levy et al.~\cite{levy} 
	   to the H1 LRG data (a) and to the ZEUS $M_X$ data (b).}
  \label{fig:f5}
\end{figure}

In Fig.~\ref{fig:f4} a comparison is shown between the 
diffractive PDFs extracted from the NLO QCD fit by P. Newman et al. 
to the ZEUS $M_X$ data 
(solid line) and from the same fit to the H1 LRG data (shaded line), 
the latter being essentially the well known H1 fit 2002~\cite{h1lrg}. 
Note that most of the 
data points from the high $\beta$ region, where discrepancies arise between 
the data sets (Fig.~\ref{fig:f2}), have not been included in the fit. 
As a reflection of the difference in the scaling violations between the two 
sets of measurements (Fig.~\ref{fig:f2}), 
the quark density is similar at low $Q^2$ and 
evolves differently to higher $Q^2$; the gluon density is a factor $\sim 2$ 
smaller in the ZEUS data than in the H1 sample. This disagreement is confirmed 
and quantified in Fig.~\ref{fig:f5}, which shows the fraction of the Pomeron 
momentum carried by quarks (red/dark line) and by gluons (blue/light line), 
as a function of $Q^2$, 
as resulting from the fit by A. Levy et al.; this fit is similar to that 
of P.~Newman et al., but completely independent, performed on the H1 
LRG data (Fig.~\ref{fig:f5}a) and on the 
ZEUS $M_X$ data (Fig.~\ref{fig:f5}b). 
The fraction of the Pomeron momentum carried by gluons is 
between $70\%$ and $90\%$ in the H1 data and between $55\%$ and 
$65\%$ in the ZEUS $M_X$ data. The same study has been carried out also 
on the ZEUS proton tagged data and the resulting integral of the fractional 
momentum is in agreement with the H1 value. 

\begin{figure}
  \includegraphics[height=.25\textheight]{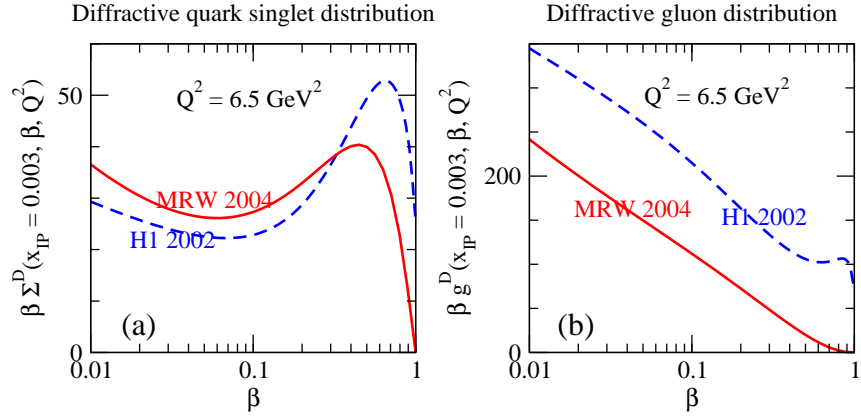}
  \put(-280,30){(a)}
  \put(-115,30){(b)}
  \caption{The diffractive parton densities resulting from a combined QCD 
           fit by A. Martin et al.~\cite{martin} 
	   to the ZEUS proton tagged, ZEUS $M_X$ and H1 LRG data. 
	   The dashed lines 
           are the densities obtained in the H1 fit 2002~\cite{h1lrg}.}
  \label{fig:f6}
\end{figure}

The same data have also been analysed according to a new approach by A. Martin 
et al.~\cite{martin}, which does not 
assume Regge factorization and shows that the collinear factorization theorem, 
though valid asymptotically in diffractive DIS, has important 
modifications at the energies 
relevant at HERA, which can be quantified using perturbative QCD. 
%According to a new approach to the analysis of the diffractive structure 
%function data~/cite{martin}, though collinear factorization theorem holds 
%in diffractive DIS, important modifications exist at the HERA energies, 
%which can be quantified using perturbative QCD. 
The DPDFs are shown to satisfy an inhomogeneous evolution equation and 
the need of including both the gluonic and sea-quark components of 
the perturbative Pomeron is considered.   
The DPDFs resulting from a combined fit to the ZEUS proton tagged data 
and $M_X$ data and to the H1 LRG data are shown in Fig.~\ref{fig:f6} 
(solid line), together with H1 fit 2002 (dashed line). 
While the quark densities are not very different from those of H1, the gluon 
distribution is significantly lower than that from H1. 

\vspace{0.2cm}

The discrepancies between the various DPDFs shown in 
Figs.~\ref{fig:f4} and~\ref{fig:f6} are large and 
presently not understood. They 
are due to a combination of effects: disagreement in the data, different 
fit methods and assumptions behind them. Therefore these differences 
between the DPDFs are at the moment the only 
estimate we have of their uncertainties. 
A precise and consistent determination of the DPDFs is certainly one of the 
main tasks that the HERA community has to face in the near future. 
Among other reasons, they are a crucial input for the prediction 
of any inclusive diffractive cross section at the LHC. 

\subsection{QCD factorization tests}

\begin{figure}
  \includegraphics[height=.20\textheight]{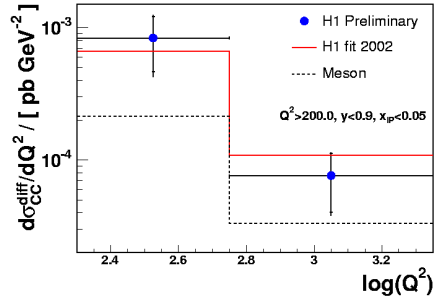}
%  \includegraphics[height=.40\textheight]{cc_ratio.eps}
%  \put(-350,2){(a)}
%  \put(-130,2){(b)}
  \caption{H1 charge current differential cross section 
           $d\sigma^{diff}_{cc}/dQ^2$ as a function of $log(Q^2)$.}
  \label{fig:f7}
\end{figure}

According to the factorization theorem, calculations based on DPDFs 
extracted from inclusive measurements should allow to predict cross 
sections for 
other diffractive processes. Calculations based on H1 fit 2002 agree well 
with the data on 
diffractive $D^*$ production in DIS~\cite{h1dstardis} and 
diffractive dijet production in DIS~\cite{h1dijetdis}. A further test  
of factorization comes from the study of events with a large rapidity 
gap in charged current interactions at high $Q^2$: in Fig.~\ref{fig:f7}   
the differential cross section $d\sigma^{diff}_{cc}/dQ^2$, as measured by 
H1~\cite{h1chargecurrent}, is presented as a function of $Q^2$ and is 
well described by a calculation based on H1 fit 2002. A similar result 
was obtained by ZEUS~\cite{zeuschargecurrent}. 
However, 
the important uncertainties on the DPDFs discussed in the previous section 
make the conclusions on the validity of QCD factorization in DIS rather 
weak. 

The factorization theorem does not hold in the case of 
diffractive hadron-hadron scattering~\cite{fact}: 
indeed it has been known for years that 
the DPDFs extracted from HERA data overestimate the rate of 
diffractive dijets at the TEVATRON by one order of 
magnitude~\cite{tev_facto_break}. 
It was shown in~\cite{GSP} that this breakdown of factorization
can be explained by screening (unitarization) effects.
In the $t$-channel Reggeon framework, these
effects are described by multi-Pomeron exchange diagrams.
Because of the screening, the probability of rapidity
gaps in high energy interactions to survive decreases since
they may be populated by rescattering processes.
The screening corrections are accounted for by the introduction
of a suppression factor, which is often called
the {\it survival probability of rapidity gaps}.
As shown in~\cite{GSP} and ~\cite{kkmr2}, the current CDF
diffractive dijet data, with one or two rapidity gaps,
are in good quantitative
agreement with the multi-Pomeron-exchange model.

In photoproduction at HERA ($Q^2 \sim0$), the exchanged photon, which is real 
or quasi real, can either interact 
directly with the proton or first dissolve into partonic constituents 
which then scatter off the target (resolved process). 
In the former case dijet photoproduction is described by a photon gluon 
fusion process. In the latter case the photon behaves 
like a hadron. 
Factorization should then be valid for direct interactions as in the case 
of DIS with large $Q^2$, whereas for the resolved contribution it is 
expected to fail due to rescattering corrections.
In the ideal theoretical limit, the suppression factor of 0.34 is 
evaluated 
for the resolved process within the multi-Pomeron exchange 
model~\cite{supp_gammaprod}. However, in reality there is no clear
model independent separation between
the direct and resolved processes. In particular, the direct contribution 
is smeared by the experimental resolution and uncertainties. Moreover, at 
NLO these contributions are
closely related. Recently Klasen and Kramer~\cite{kk} have performed an 
analysis of diffractive dijet photoproduction data at NLO where they 
suppressed the resolved process by a factor~0.34.

\begin{figure}
  \includegraphics[height=.30\textheight]{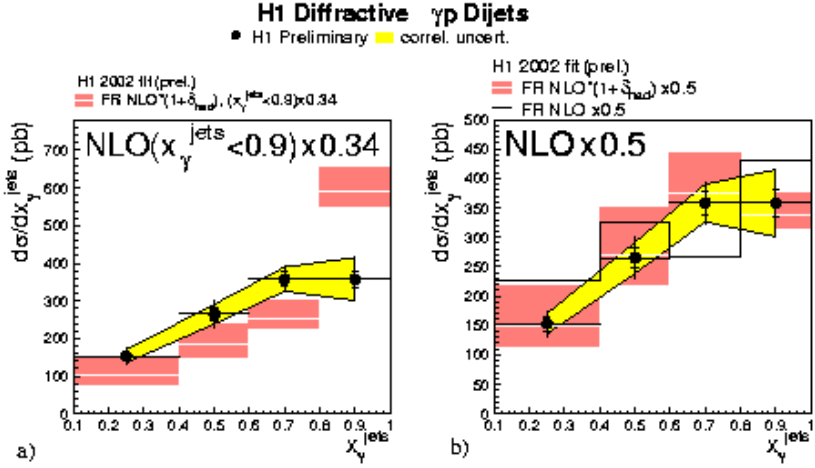}
  \caption{H1 cross section for the diffractive production of dijets  
           in photoproduction as a function of $x_\gamma$. In (a) only 
	   the resolved contribution to the NLO calculation 
           has been scaled by the factor 0.34, while in 
	   (b) the complete prediction is multiplied by a factor 0.5.}
  \label{fig:f8}
\end{figure}

\begin{figure}
  \includegraphics[height=.25\textheight,angle=-90]{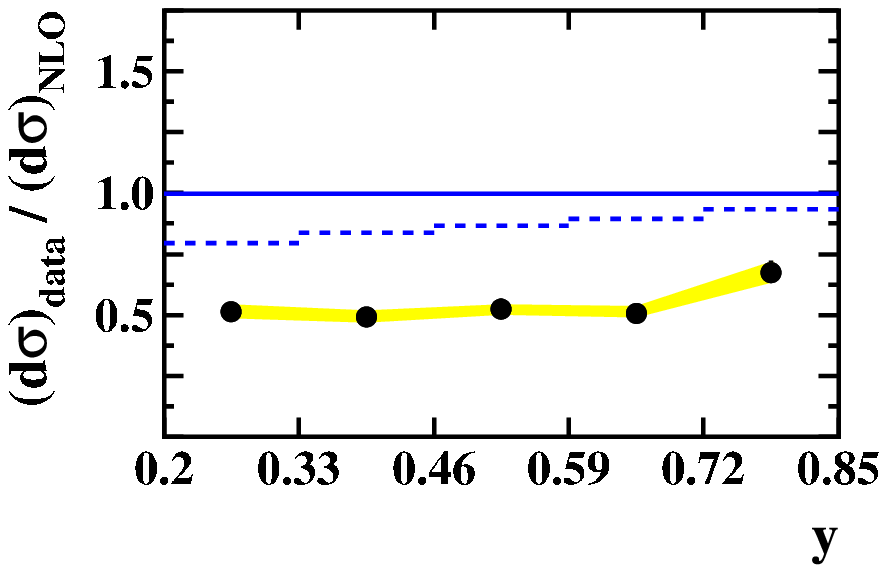}
  \includegraphics[height=.25\textheight,angle=-90]{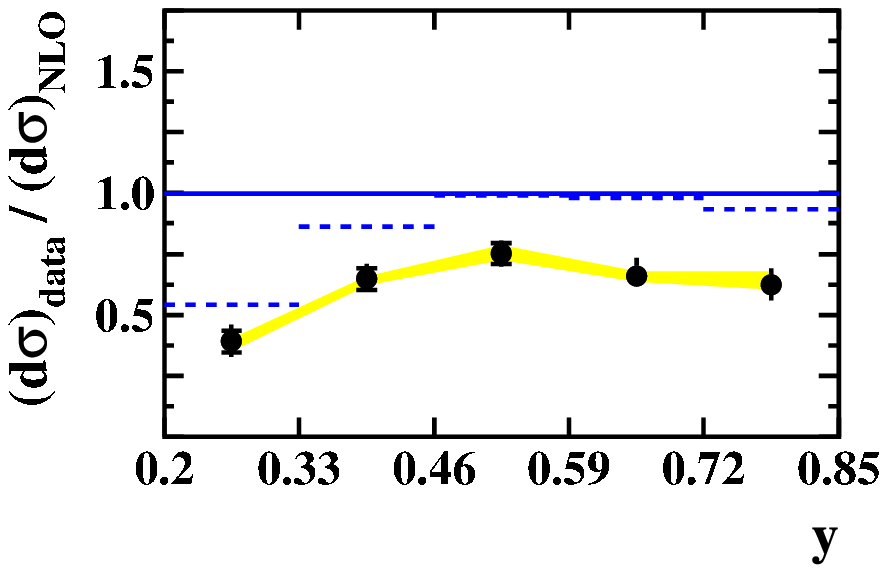}
  \includegraphics[width=.21\textwidth,angle=-90]{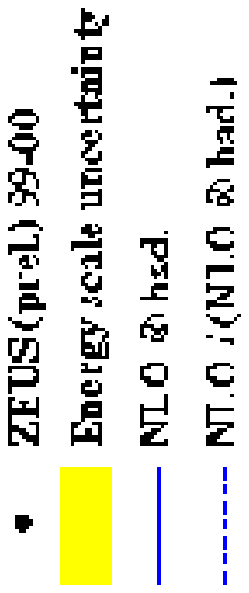}
  \put(-370,-19){$x_{\gamma}^{\rm{obs}}>0.75$}
  \put(-210,-19){$x_{\gamma}^{\rm{obs}}<0.75$}
  \put(-370,-99){(a)}
  \put(-210,-99){(b)}
  \caption{Ratio of the ZEUS diffractive dijet data to the NLO QCD 
           predictions~\cite{kk} of the single differential cross section in 
	   $y$ for the sample enriched in direct (a) and resolved 
	   (b) photoproduction.}
  \label{fig:f9}
\end{figure}

Fig.~\ref{fig:f8} shows the differential cross section, as measured 
by H1~\cite{h1dijetdis}, 
for the diffractive photoproduction of two jets as a function of 
$x_\gamma$ (the fraction of the photon 
momentum entering the hard scattering), where the NLO prediction 
has been tested in two different weighting schemes: 
in Fig.~\ref{fig:f8}a only the resolved part has been scaled by the factor 
0.34, while in 
Fig.~\ref{fig:f8}b a global suppression factor 0.5 is applied to both 
the direct and resolved components. 
In Fig.~\ref{fig:f9} the ratio of the ZEUS data~\cite{zeusdijet} to the 
NLO predictions of Klasen and Kramer~\cite{kk} with no suppression factor 
($R=1$) 
is shown separately for the sample enriched (a) in the direct 
($x_\gamma$ $\ge 0.75$) and (b) in the resolved ($x_\gamma$ < 0.75) 
components.  Both for resolved 
and direct photoproduction the ratio is flat, but the data are lower by a 
factor of $\sim 2$ compared to the NLO calculations. 
The overall message from the data of Figs.~\ref{fig:f8} and~\ref{fig:f9} is 
that, while a suppression of only the resolved contribution at NLO is 
disfavored by the data, a good agreement is achieved with the global 
suppression 0.5, which furthermore yields a good description of all measured 
cross sections.

The fact that the data, 
apparently against expectations, support suppression of 
direct photoproduction, has been addressed by M.~Klasen~\cite{klasen} 
and has been related to the critical role of an initial state singularity in 
the way factorization breaks down and to the need of a modification of the 
suppression mechanism:
%Modification of suppression mechanism, new factorization scheme. 
separation of direct and resolved photoproduction events is a 
leading order concept. At NLO they are closely connected. 
The sum of both cross sections is the only physical relevant observable, 
which is approximately independent of the factorization scale, 
$M_\gamma$~\cite{trediciklasen}. 
By multiplying the resolved cross section with the suppression factor 
$R=0.34$, the scale dependence of the NLO direct cross section is compensated 
against that of the LO resolved part~\cite{kk}. 
But at NLO collinear singularities arise from the photon initial state,
 which are absorbed at the factorization scale into the photon PDFs; 
the latter become in turn $M_\gamma$ dependent. 
An equivalent $M_\gamma$ dependence, just with the opposite sign, is then 
left in the NLO corrections to the direct contribution. 
Hence, in order to get a physical cross section at NLO, that is the 
superimposition of the NLO direct and LO resolved cross section, and to 
restore the scale invariance, one must multiply the $M_\gamma$ dependent 
term of the NLO correction to the direct contribution with the same 
suppression factor as the resolved cross section. 

\vspace{0.2cm}

The situation with factorization breaking in dijet 
photoproduction is not completely understood and further
experimental and theoretical efforts are needed. On the other hand the 
uncertainties on the DPDFs largely discussed previously are a further ingredient 
which makes it difficult a clear understanding. 
As was emphasized in~\cite{supp_gammaprod}, a possible way to study the 
effects of factorization breaking due to rescattering in diffractive  
photoproduction is to measure the ratio of diffractive and inclusive dijet 
photoproduction as a function of $x_\gamma$. For such a quantity (at least) 
some of the theoretical and experimental uncertainties will cancel.

The understanding of factorization breaking 
in hadron-hadron collisions is of fundamental importance 
for the diffractive physics at the LHC. 
The rapidity gap survival factor is an essential ingredient of the 
predictions~\cite{Khoze:2000} on exclusive diffractive 
Higgs production, which will be discussed in 
the last section. 
%This is why understanding the HERA diffractive 
%dijet photoproduction data is a crucial issue. 

\subsection{Diffraction at the TEVATRON}

As discussed in the previous section, factorization is not expected to hold 
in hadron-hadron collisions. A strong breakdown of factorization at the 
TEVATRON has been known for some time from run-I (1992-1995) 
results~\cite{dinodis05}: the 
single-diffractive to non-diffractive 
ratios for dijets, $W$, $b$-quark and $J/\psi$ production, as well as the 
ratio of double-diffractive to non-diffractive dijet production are all 
$\sim 1\%$, a factor 10 less than at HERA.  
%It is, however, interesting to note that, except for the 
%overall suppression in normalization, factorization approximately hold at 
%fixed $\sqrt{s}$. Furthermore, the ratios of two-gap to one-gap cross 
%sections for both soft and hard processes appear to obey factorization.  
However, the ratio of double- to single-diffractive dijets is found to be 
about a factor 5 larger than the ratio of single- to non-diffractive dijets, 
suggesting that there is only a small extra suppression when going from one 
to two rapidity gaps in the event, as confirmed by 
predictions~\cite{kkmr2}.   
In this respect the TEVATRON data are being a very powerful tool to shed 
light on the factorization breaking mechanism. 

\vspace{0.2cm}

One of the major challenges of run-II is the measurement of central 
exclusive production rates (dijets, $\chi_c^0$, diphotons). 
By central exclusive, we refer to the process 
$p\bar{p}\rightarrow p \oplus \phi \oplus \bar{p}$, where 
$\oplus$ denotes the absence of hadronic activity ('gap') 
between the outgoing hadrons and the decay products of the central 
system $\phi$.
As we will discuss in the last section, the exclusive Higgs signal is 
particularly clean and the signal-to-background ratio is especially favorable, in comparison with other proposed selection modes. However, the expected number 
of events is low. Therefore it is important to check the predictions for exclusive Higgs production by studying processes mediated by the same mechanism, but 
with rates which are high enough to be observed at the TEVATRON 
(as well as at the LHC)~\cite{exclusive}. 
%New data on exclusive rates are currently being analysed at TEVATRON. 

\begin{figure}
  \includegraphics[height=.25\textheight]{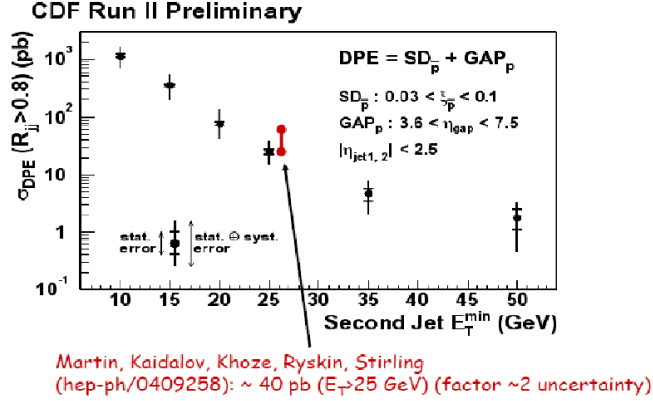}
  \caption{CDF dijet production cross section for $R_{jj}>0.8$ in double 
Pomeron exchange events as a function of $E_T^{min}$, the $E_T$ of the lower 
$E_T$ jet. }
  \label{fig:f10}
\end{figure}

The CDF search for exclusive dijet production is based on the 
reconstruction of 
the dijet mass fraction $R_{jj}$ in double Pomeron exchange events. $R_{jj}$ 
is defined as the mass of the two leading jets in an event divided by 
the total mass measured in all calorimeters. 
At first sight, we might expect that the exclusive dijets
form a narrow peak concentrated at $R_{jj}$ close to 1.
In reality, the peak is smeared out due to hadronization
and jet searching procedure as well as due to a 'radiative
tail' phenomenon~\cite{kmrG}. So it is not so surprising that
within the CDF selection cuts no peak has been seen. CDF reports 
production cross sections for events with $R_{jj}>0.8$, which are 
interpreted as the upper limits for exclusive production. 
Fig.~\ref{fig:f10}~\cite{dinodis05} 
shows such cross 
sections as a function of $E_T^{min}$, the $E_T$ of the lower $E_T$ jet. 
These data agree, within errors, with recent predictions for exclusive dijet 
production~\cite{exclusive}. The analysis benefits from using 
dijet events in which at least one of the jets is $b$-tagged: presently 
more data on heavy flavor exclusive dijets are being collected with a 
special $b$-tagged dijet trigger. 

\vspace{-0.2cm}

\subsection{Diffraction at RHIC}

New interesting experimental results from RHIC were presented by 
Guryn, White and Klein. 
In particular, Guryn \cite{gur} described the results of the 
measurement of the single spin analyzing power $A_{N}$ in polarized pp 
elastic scattering at 200 GeV.
The recent results on inelastic diffraction with Au-Au, d-Au and pp beams 
were reviewed by White \cite{sw}. Klein \cite{sk} showed the results of 
the STAR collaboration for coherent photonuclear $\rho$ and 4 charged pion 
production.
%$\rho0$ and $\pai+pai- pai+pai-$

\subsection{Updates on theory}

Several excellent mini review type theoretical talks were presented. 
Hard diffraction in DIS and the origin of hard Pomeron from rescattering
were discussed by Brodsky \cite{brod}. He also reviewed such  effects as
Color Transparency, Color Opaqueness and Intrinsic Charm. 
Levin \cite{lev} gave a brief review of the current status of high density 
QCD with its ups and downs. The recent progress in the BFKL studies was 
covered by Andersen \cite{and}. In particular, he discussed the high-energy 
limit of diffractive scattering processes in the BFKL resummation framework.
He showed  that the BFKL equation was solved at full next-to-leading 
logarithmic accuracy.

%%%%%%%%%%%%%%%%%%%%%%%%%%%%%%%%%%%%%%%%%%%%%%%%%%%%%%%%%%%%%%%%%%%%%%%%%%%%%%

\section{Exclusive Meson Production and DVCS}

%\subsection{Vector meson production}

The dynamics of diffractive interactions can also be studied through 
exclusive vector meson ($ V = \rhoz, \omega, \jpsi, ... $) 
and photon production, $ l^{\pm} + N \longrightarrow l^{\pm} + V + Y$, 
where $Y$ is either an elastically scattered nucleon or a low-mass
state dissociative system. At low transverse momentum transfer
at the nucleon vertex, the photoproduction of \rhoz, $\omega$ and $\phi$
mesons is characterized by a ``soft'' dependence of their cross-sections
in the $\gamma p$ center-of-mass energy, $W$. This can be interpreted 
in the framework of Regge theory as due
to the exchange of a ``soft'' Pomeron (\pom) resulting in an energy dependence
of the form $\rm{d}\sigma / \rm{d} t \ \propto \ W^{4(\apom(t)-1)}$, 
where the Pomeron trajectory is parametrised as 
$\apom(t) = \apom(0) + \aprim t \simeq 1.08 + 0.25 t$. 
However, in the presence of a ``hard'' scale like large values of the photon 
virtuality \qsq or of the momentum transfer \ttra\ 
or of the vector meson mass, perturbative QCD (pQCD) is expected to
apply. Diffractive vector meson production can then be seen in the nucleon
rest frame as a sequence of tree subprocesses well separated in time:
the fluctuation of the exchanged photon in a \qqbar\ pair, the hard 
interaction of the \qqbar\ pair with the nucleon via the exchange of
(at least) two gluons in a color singlet state, and the \qqbar\ pair
recombination into a real vector meson. This approach results in a stronger
rise of the cross section with $W$, which reflect 
the strong rise at small $x$ of the gluon density in the nucleon. 
Such an energy dependence is observed in \jpsi\ production, where the
quark charm mass provides a hard scale. It is of particular interest
to study the role of other hard scales like \qsq\ and $t$ 
as well as the transition from a ``soft'' to ``hard'' behavior expected
for light vector mesons. Furthermore, to take into account the  skewing effect, i.e. 
the difference between the proton momentum fractions carried by the two 
exchanged gluons, one has to consider generalized parton distributions 
(GPDs).
GPDs are an extension of standard PDFs, which include additional information
on the correlations between partons and their transverse motion.
There are four different types of GPDs, $H(x,x^{\prime},t)$ and
$E(x,x^{\prime},t)$, where $x$ and $x^{\prime}$ are the momentum fraction
of the two parton considered, in the unpolarized case to which one should
add $\tilde{H}(x,x^{\prime},t)$ and $\tilde{E}(x,x^{\prime},t)$ in
the polarized case. While $E$ and $\tilde{E}$ have no equivalent in the
ordinary PDFs approach, $H$ and $\tilde{H}$ reduce to the usual unpolarized 
and polarized PDFs respectively in the forward limit ($x=x^{\prime}$ and $t=0$).

\begin{figure}
  \includegraphics[height=.25\textheight]{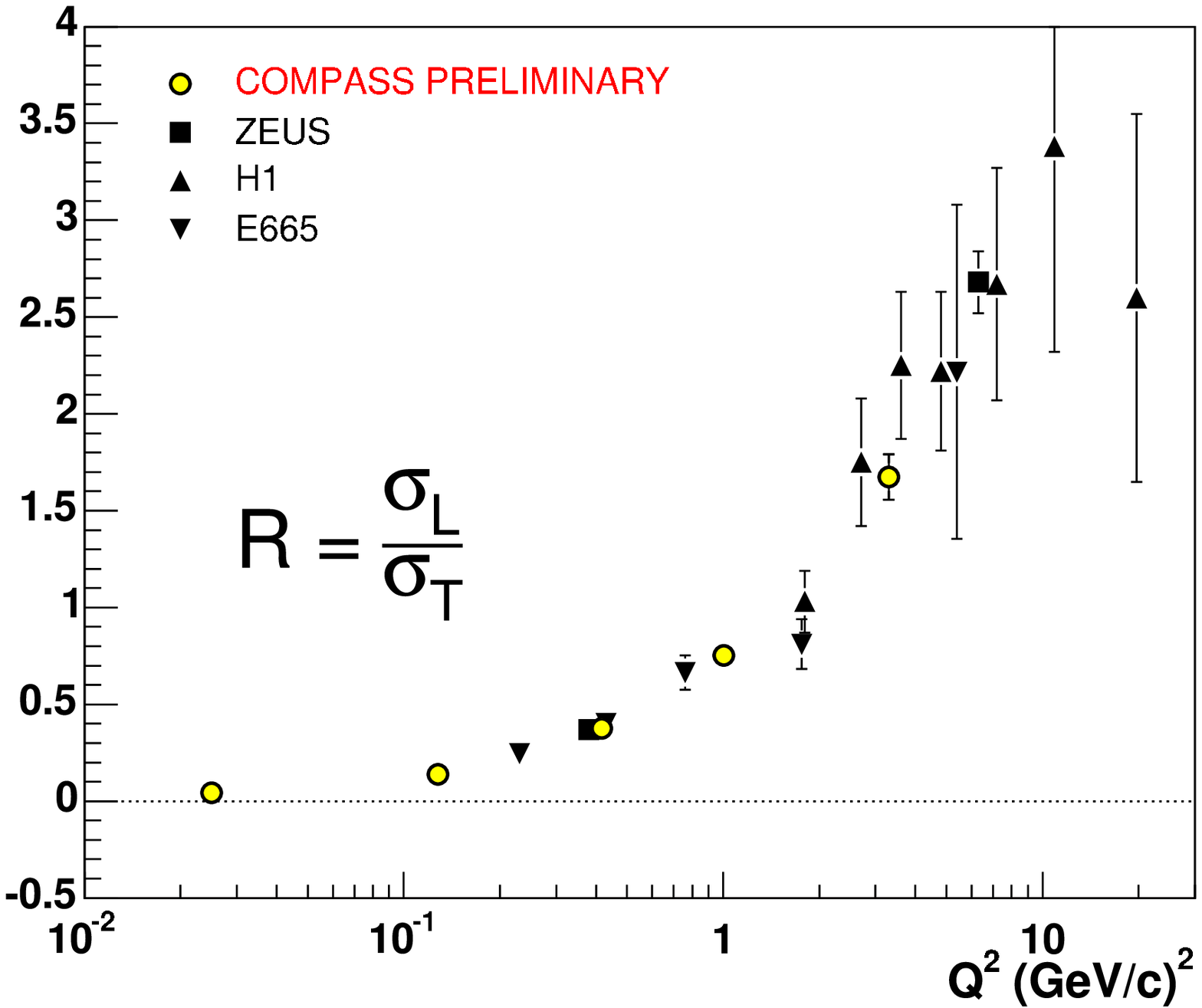}
  \includegraphics[width=.5\textwidth,height=.25\textheight]{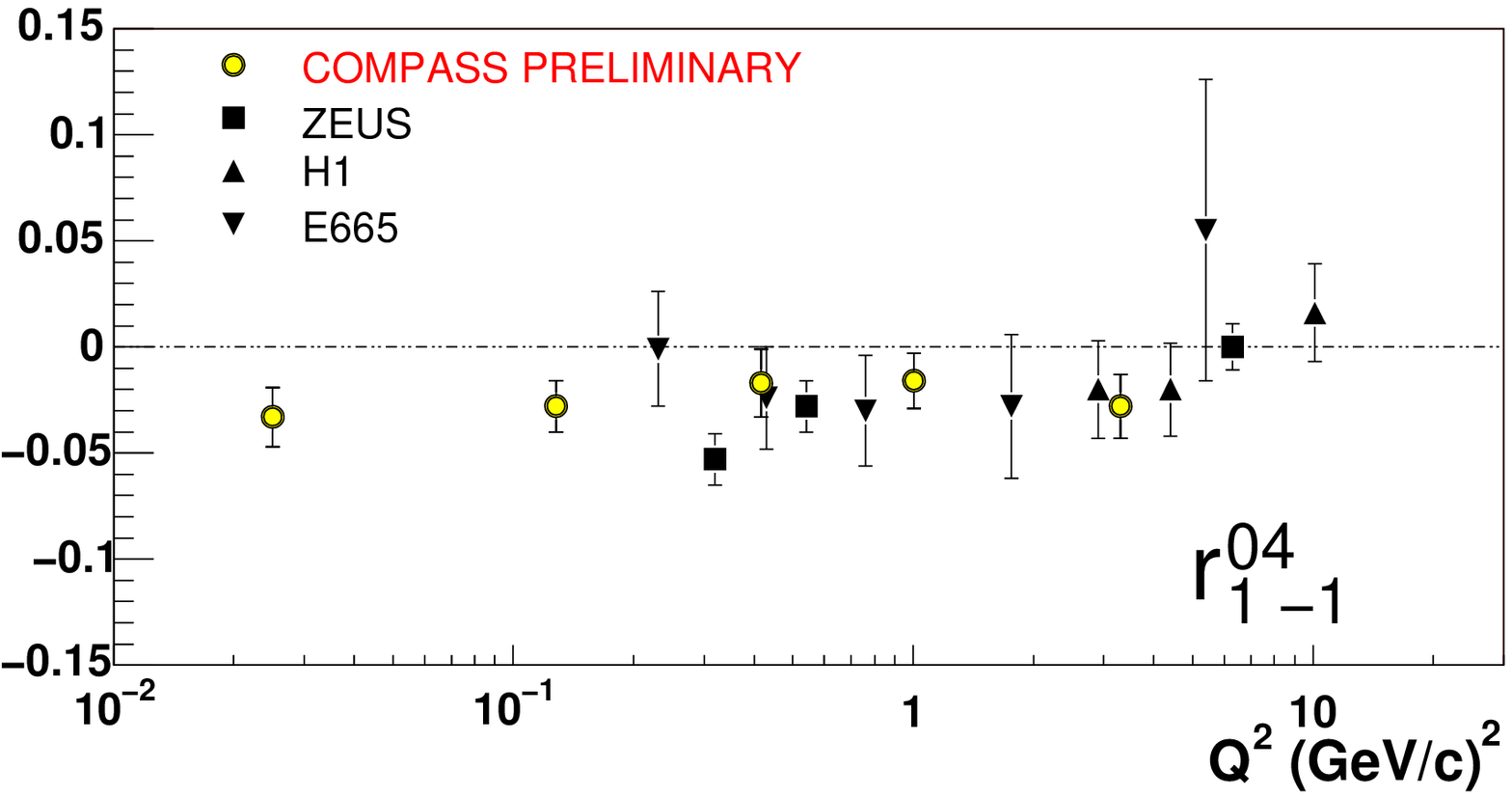}
  \put(-250,43){(a)}
  \put(-180,43){(b)}
  \caption{\qsq\ dependence (a) of the ratio $R$ between the longitudinal 
           ($\sigma_L$) and the transverse ($\sigma_T$) cross sections 
           and (b) of the \rzqzz\ matrix element
           for elastic leptoproduction of \rhoz\ as measured by COMPASS. }
  \label{fig:rhocompass}
\end{figure}

The COMPASS experiment has presented~\cite{compass:rho} a study of the 
diffractive elastic leptoproduction of \rhoz\ mesons, 
$\mu + N \longrightarrow \mu + \rhoz + N$, where $N$ is a quasi-free
nucleon from any of the nuclei of their polarized target, at 
$<W> = 10$ \gev\ for a wide range of \qsq, $0.01 < \qsq < 10$ \gevsq .
Several spin density matrix elements (SDME), which carry information
on the helicity structure of the production amplitudes, have been
extracted from the production and decay \rhoz\ angular distributions. 
The COMPASS data provide a large statistics which allows to extend 
the previous measurements towards low \qsq.  Measurements of the
\rzqzz\ matrix element, which can be interpreted as the fraction
of longitudinal \rhoz\ in the sample, have been performed as a function
of \qsq. If one assumes $s$-channel helicity conservation (SCHC) between
the exchanged photon and the \rhoz\ meson, one can obtain the  ratio $R$ 
between the longitudinal ($\sigma_L$) and the transverse ($\sigma_T$) 
cross sections (see Fig.~\ref{fig:rhocompass}a). A weak violation of SCHC
is observed through the \rzqumu\ matrix element (see Fig.~\ref{fig:rhocompass}b, in agreement with results of previous experiments. It has to be noted
that the study of systematic effects is still ongoing and that only
the statistical errors are provided. 

Elastic electroproduction of $\phi$ mesons has been studied in $e^{\pm}p$
collisions by the ZEUS experiment~\cite{zeus:phi} in the kinematic range
$2 < \qsq < 70 $ \gevsq, $35 < W < 145$ \gev\ and $\ttra < 0.6$ \gevsq. 
The energy dependence of the $\gamma^* p$ cross section has been 
measured and can be parametrised
as $\sigma \propto W^{\delta}$, with $\delta \simeq 0.4$. This value is 
between the ``soft'' diffraction value and the one observed for \jpsi.
No \qsq\ or $t$ dependence of the slope $\delta$ was observed with the 
present precision. When parametrised as a falling exponential,
the $t$ dependence of the cross section leads to $b$ slopes
in the range from $6.4 \pm 0.4$ \gevsqinv\ at $\qsq = 2.4$ \gevsq\ to 
$5.1 \pm 1.1$ \gevsqinv\ at $\qsq = 19.7$ \gevsq. The values of $\delta$
and $b$ were found to scale with respect to other vector mesons results
when plotted as a function of $\qsq + \mvm^2$, where \mvm\ is the mass
of the vector meson, suggesting that this could be a good approximation 
of the universal scale in this process. The ratio between the longitudinal 
($\sigma_L$) and the transverse ($\sigma_T$) cross sections, extracted from
the $\phi$  angular distributions, was found to increase with \qsq\ and
when compared with results obtained for other vector mesons to scale
with $\qsq / \mvm^2$. 

\begin{figure}
  \includegraphics[height=.28\textheight]{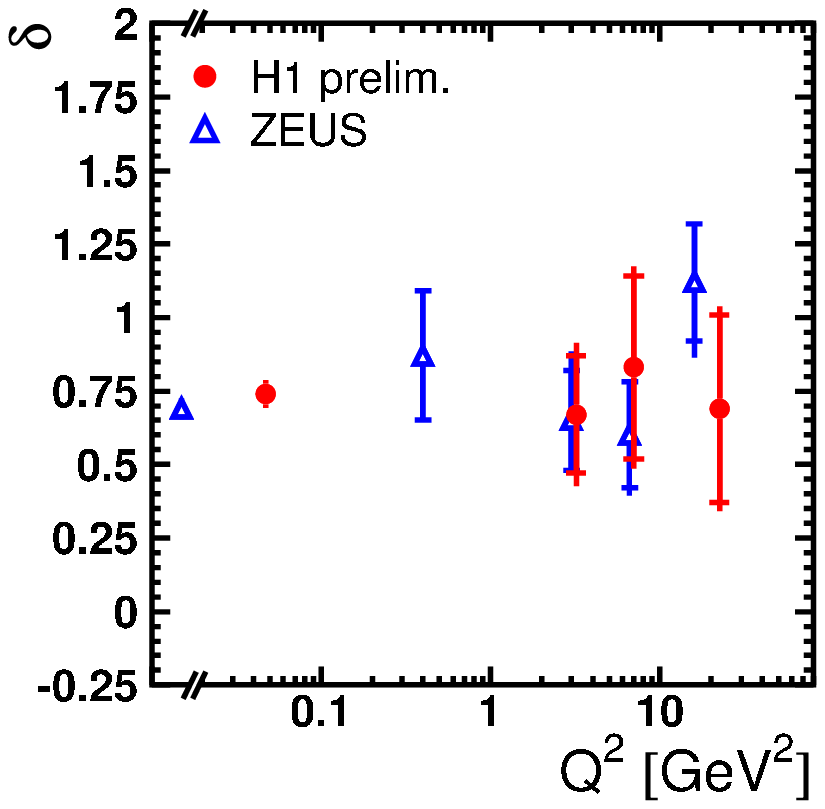}
  \includegraphics[height=.27\textheight]{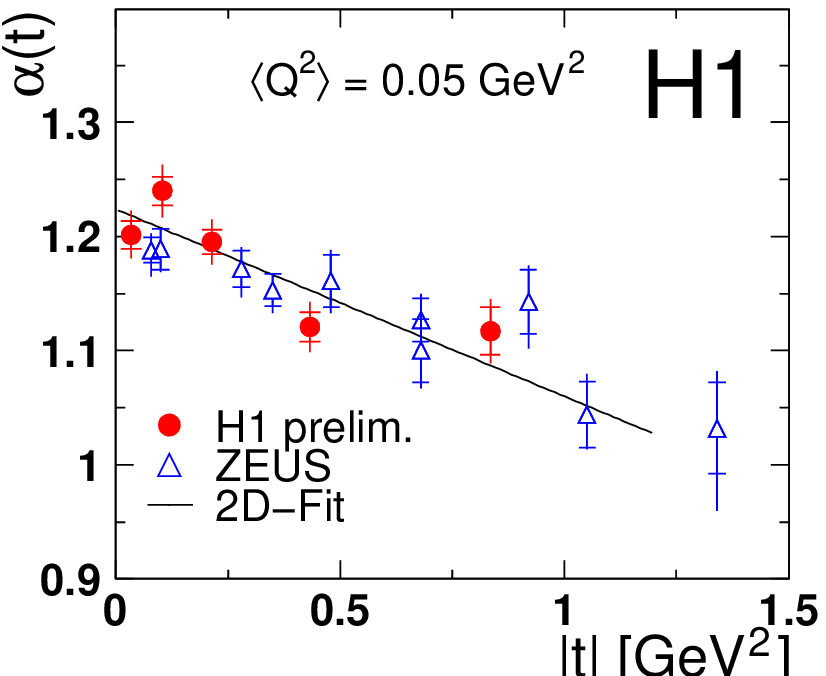}
  \put(-250,33){(a)}
  \put(-50,33){(b)}
  \caption{(a) The $\sigma \propto W^{\delta}$ fit parameter $\delta$ 
               for \jpsi\ production as a function of \qsq.
           (b) The effective trajectory $\apom(t)$ as a function of $t$
               for \jpsi\ photoproduction.} 
  \label{fig:jpsih1}
\end{figure}

H1 has presented~\cite{h1:jpsi} comprehensive results on elastic \jpsi\
production in the $\gamma^{*}p$ center-of-mass energy ranges 
$40 < W < 305 $ \gev\ in photoproduction and
$40 < W < 160 $ \gev\ in electroproduction up to 80 \gevsq\ in \qsq\
and in both cases for $\ttra < 1.2$ \gevsq. In such a process, the hard
scale provided by the mass of the involved charm quark ensures the validity
of a pQCD description. This is even more so in electroproduction where \qsq\
can provide a second hard scale. The \qsq\ and $W$ dependent $\gamma^{*}p$
cross-sections have been extracted. A steep rise with energy, $\sigma \propto
W^{\delta}$, was observed with values of $\delta \simeq 0.7$ independently
of \qsq\ (see Fig.~\ref{fig:jpsih1}a). The effective Pomeron trajectories
$\apom(t) = \apom(0) + \aprim t$ have been extracted from the study of
the doubly differential $\rm{d}\sigma/\rm{d}t$ cross-section as a function
of $W$ and $t$. In photoproduction (see Fig.~\ref{fig:jpsih1}b), a
positive value of $\aprim = 0.164 \pm 0.028 \pm 0.030$ \gevsqinv\ was
obtained, leading to a shrinkage of the forward scattering peak, even if the
effect is smaller than observed in hadron-hadron interactions.
In electroproduction, within its large error, the obtained value of \aprim\
was found compatible both with the photoproduction result and zero.
Finally, the helicity structure has been analyzed as a function of \qsq\ and
$t$ and no evidence for a violation of SCHC has been observed. Assuming SCHC,
the ratio of the longitudinal and the transverse cross sections has been
extracted as a function of \qsq.

\begin{figure}
  \includegraphics[height=.25\textheight]{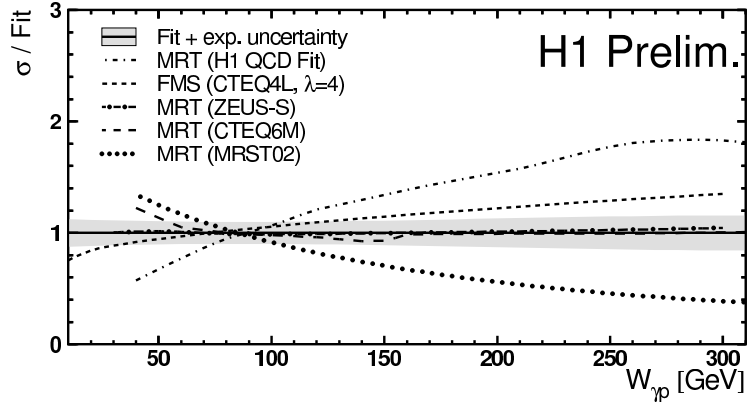}
  \caption{Ratio versus $W$ of the MRT~\cite{teubner} theoretical predictions
           based on several gluon distributions to a parametrization
           of H1 \jpsi\ photoproduction preliminary data.}

  \label{fig:teubner}
\end{figure}

Teubner~\cite{teubner} has presented a model for vector meson production
based on $k_T$ factorization, which uses a parton-hadron duality ansatz to
avoid the large uncertainties arising from the poorly known vector meson
wave functions. The predictions obtained for \jpsi\ cross section
as a function of $W$ (see Fig.~\ref{fig:teubner}) with different sets of gluon 
distribution show a huge spread. This indicates a
possible sensitivity to the gluon at small $x$ and small to intermediate
scales, i.e. a kinematic region where fits to the inclusive data do not
constrain the gluon with high precision. Getting high precision data on
vector meson production at HERA and reducing the remaining theoretical 
uncertainties might then allow to pin down the gluon at low $x$.

Kroll \cite{kroll} presented a LO QCD calculation for light vector meson
electroproduction taking into account the transverse momenta of the quark
and the anti-quark as well as Sudakov factors. The GPDs are modeled
according to the ansatz of Radyushkin and Gaussian wavefunctions are used
for the vector mesons. A fair agreement with the available data on \rhoz\
and $\phi$ production at HERA is obtained between the predictions
for the transverse and the longitudinal cross section as well as for
the spin density matrix elements.

\begin{figure}
  \includegraphics[height=.32\textheight]{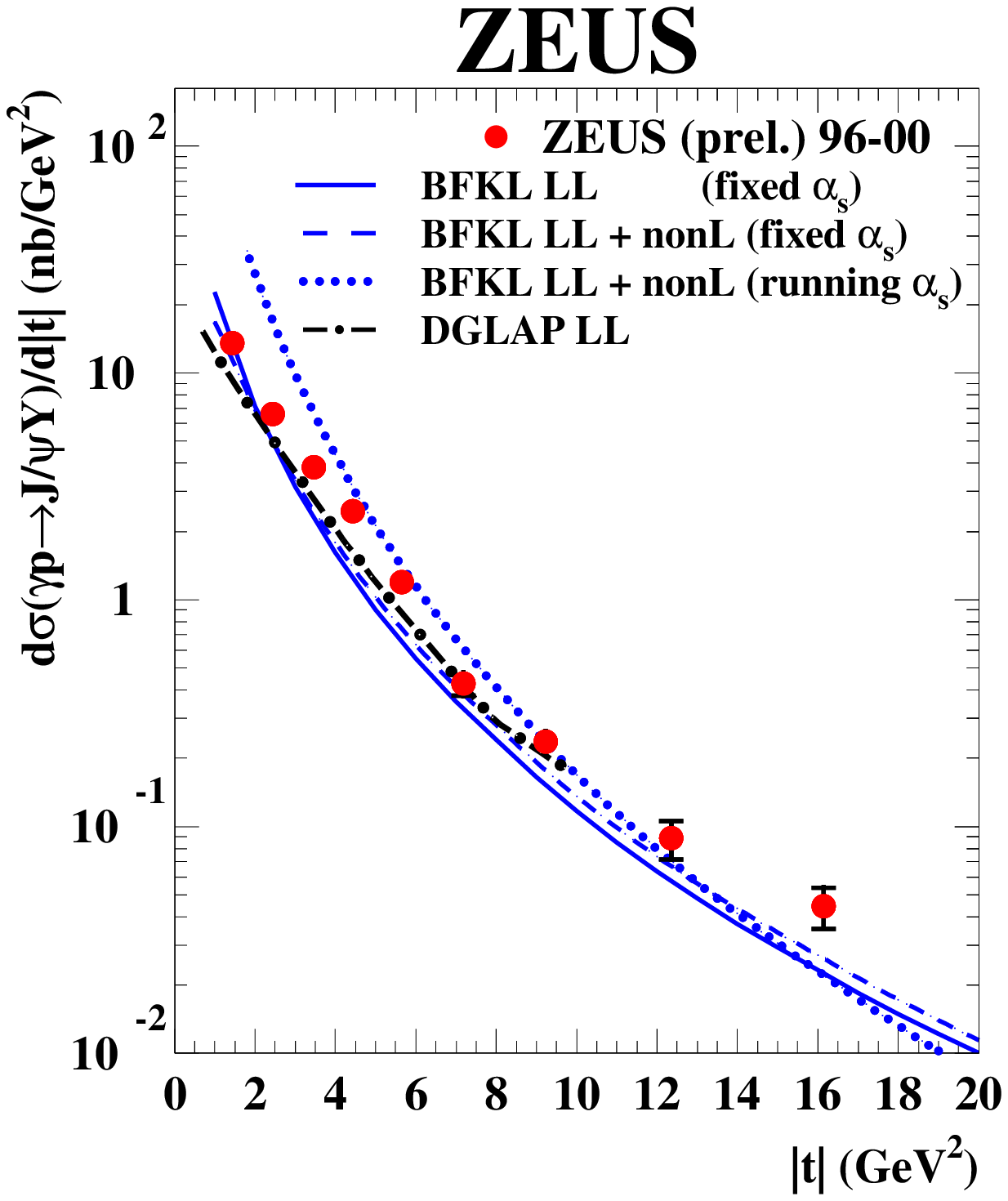}
  \includegraphics[height=.32\textheight]{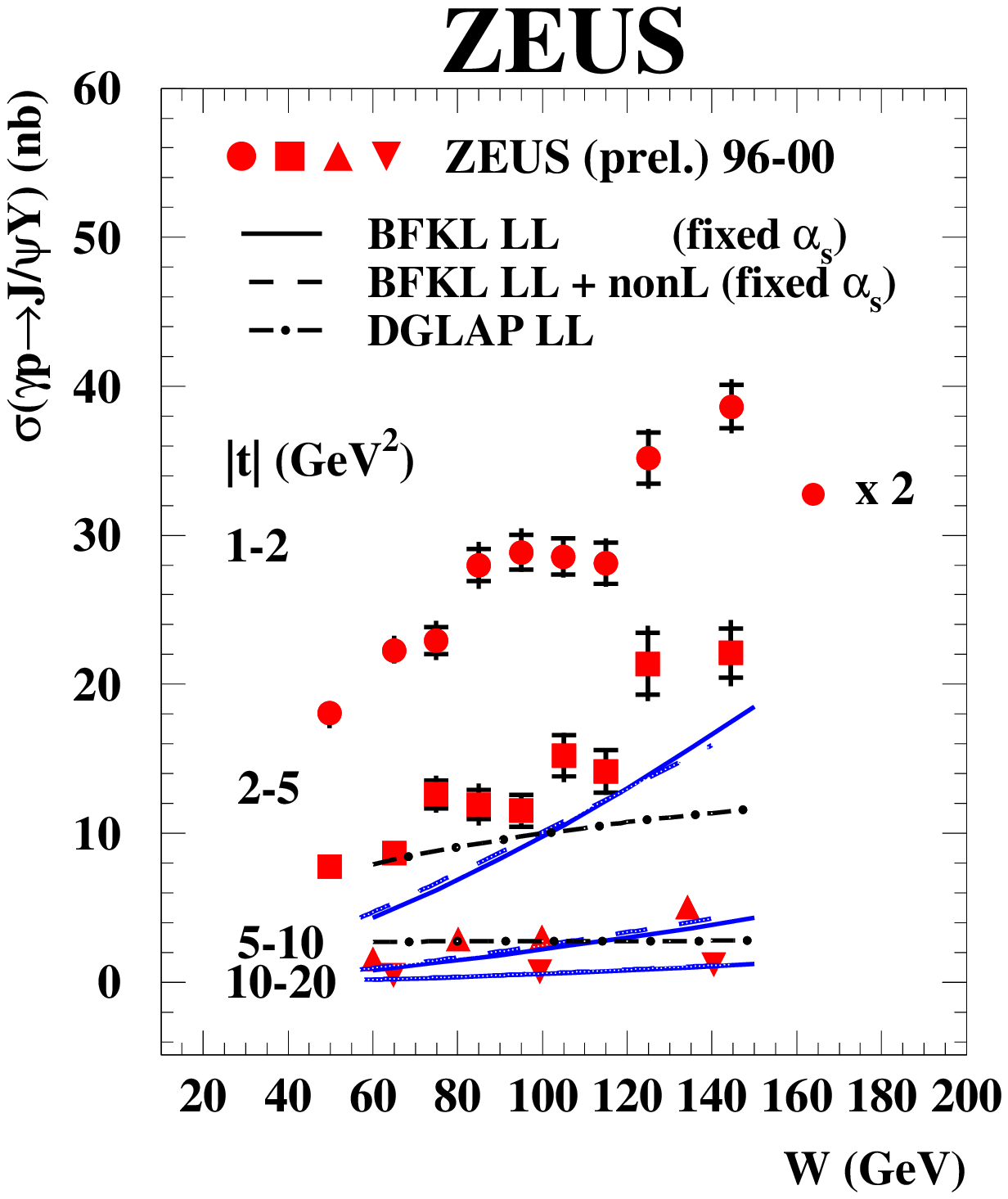}
  \put(-310,37){(a)}
  \put(-35,37){(b)}
  \caption{The proton-dissociative diffractive photoproduction of J/psi mesons 
           cross sections for $\ttra >1$ \gevsq\ (a) as a function of $t$ and
           (b) as a function of $W$ for four bins in $t$ compared to 
           predictions from  based on BFKL and DGLAP.
           }
  \label{fig:vmht}
\end{figure}

Photoproduction of vector mesons at large \ttra\ is largely studied since
a few years as it is expected to be described by perturbative models involving
the BFKL dynamics in the exchanged gluon ladder~\cite{forshaw}. These
models predict a power law behavior of the $t$ dependence of the
cross section and a rise with \ttra\ of the steepness of the $W$
dependence.

H1 has presented~\cite{h1:rhoht} results on \rhoz\ photoproduction in the
kinematic range $ 75 < W < 95$ \gev\ and $1.5 < \ttra < 10 $ \gevsq\ where
the mass of the proton dissociative system $Y$ is limited to $M_Y < 5$
\gev. The measured $t$ dependence of the cross-section is well described
by a power law of the form $\ttra^{-n}$ with $n = 4.41 \pm 0.07 
^{+0.07}_{-0.10}$ and can be reproduced by BFKL model predictions.
A study of the helicity structure has been performed and confirms the
violation of SCHC in the case of \rhoz\ photoproduction at large \ttra\, 
in contrast to what was observed for high \ttra\ \jpsi\ production
\cite{h1:jpsiht,zeus:vmht}. This is generally attributed to differences
in the wave function between $\rho$ and \jpsi. 

\jpsi\ photoproduction at large \ttra\ has been studied by 
ZEUS~\cite{zeus:jpsiht} in the kinematic range $50 < W < 150$ \gev,
$\ttra > 1$ \gevsq\ and $M_Y < 30$ \gev. Both the $t$ dependence
and the $W$ dependence of the cross-section have been extracted, 
as shown on Fig.~\ref{fig:vmht}. Fits of the form $W^{\delta}$
to the $W$ dependence of the cross section lead to values of
$\delta \simeq 1$ with an indication for a rise of $\delta$ with \ttra\ . 
The model based on BFKL has been found to 
describe the $t$ dependence.

%\subsection{Exclusive reactions and generalized parton distribution functions}

\begin{figure}
  \includegraphics[height=.26\textheight]{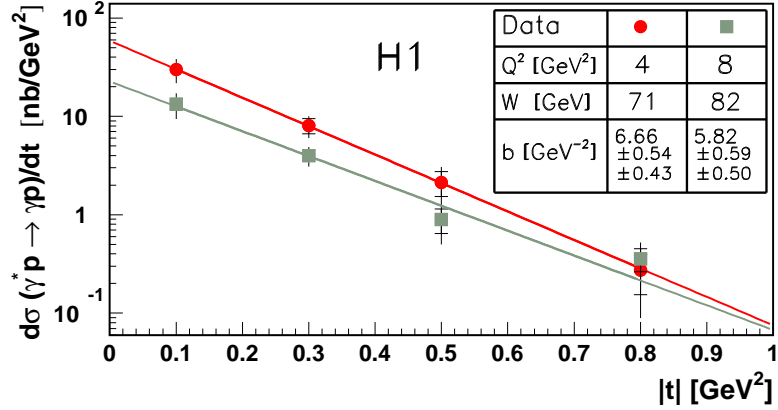}
  \caption{The DVCS cross section differential in $t$, 
           for $\qsq = 4$ \gevsq\ at $W=71$ \gev\ and 
           for $\qsq = 8$ \gevsq\ at $W=82$ \gev\, as measured by H1.
           The lines represent the results of fit of the
           form $e^{-bt}$.
           }
  \label{fig:dvcsh1}
\end{figure}

The opportunity to study Generalized Parton Distributions (GPDs)
was discussed in a common session with the Spin Physics working group.
%There are four different type of GPDs, $H(x,x^{\prime},t)$ and
%$E(x,x^{\prime},t)$, where $x$ and $,x^{\prime}$ are the momentum fraction
%of the two parton considered in the unpolarized case, to which one should
%add $\tilde{H}(x,x^{\prime},t)$ and $\tilde{E}(x,x^{\prime},t)$ in
%the polarized case. While $E$ and $\tilde{E}$ have no equivalent in the
%ordinary PDFs approach, $H$ and $\tilde{H}$ reduce to the usual unpolarized 
%and polarized PDFs in the forward limit ($x=x^{\prime}$ and $t=0$).
Information about GPDs in lepton nucleon scattering can be provided 
by measurements of exclusive processes in which the nucleon remain intact.
The simplest process sensitive to GPDs is Deeply Virtual Compton
Scattering (DVCS), i.e. exclusive photon production off the proton 
$\gamma^* p \longrightarrow \gamma p$ at small \ttra\ but
large \qsq, which is calculable in perturbative QCD.
Such a final state also receives contributions from the purely
electromagnetic Bethe-Heitler process, where the photon
is radiated from the lepton. The resulting interference term
in the cross section vanishes as long as one integrates over
the azimuthal angle between the lepton and the hadron plane. 
It is then possible to extract the DVCS cross section by
subtracting the Bethe-Heitler contribution, as done by H1 and ZEUS.
The azimuthal asymmetries resulting from the interference are
also sensitive to GPDs and are studied by HERMES, COMPASS and
at JLAB. Extracting GPDs from the DVCS process would allow,
through the Ji's sum rule,
to determine the total angular momentum carried by the quarks
which contribute to the proton spin.

A new high statistics analysis of DVCS has been performed by the
H1 experiment~\cite{h1:dvcs} in the kinematic region $2<\qsq<80$
\gevsq, $30<W<140$ \gev\ and $\ttra<1$ \gevsq. The $\gamma^* p 
\longrightarrow \gamma p$ cross section has been measured as a
function of \qsq\ and as a function of $W$. The $W$ dependence
can be parametrised as $\sigma \propto W^{\delta}$, yielding
$\delta = 0.77 \pm 0.23 \pm 0.19$ at $\qsq=8$ \gevsq, i.e.
a value similar to \jpsi\ production indicating the presence
of a hard scattering process. For the first time, the DVCS
cross section has been measured differentially in $t$ (see
Fig. \ref{fig:dvcsh1}) and the observed fast decrease with \ttra\
can be described by the form $e^{-b\ttra}$ with
$b=6.02 \pm 0.35 \pm 0.39$ \gevsq\ at $\qsq=8$ \gevsq.
This measurement allows to further constrain the models, as their
normalization depends directly on the $t$ slope parameter.
NLO QCD calculations using a GPD parametrization based on the
ordinary parton distributions in the DGLAP region and where
the skewedness is dynamically generated
provide a good description of both the \qsq\ and
the $W$ dependences. 
%Color dipole model predictions also give
%a good description of the data, in particular the one based on
%a saturation model in which the dipole is evolved according
%to the DGLAP equation.

\begin{figure}
  \includegraphics[height=.32\textheight]{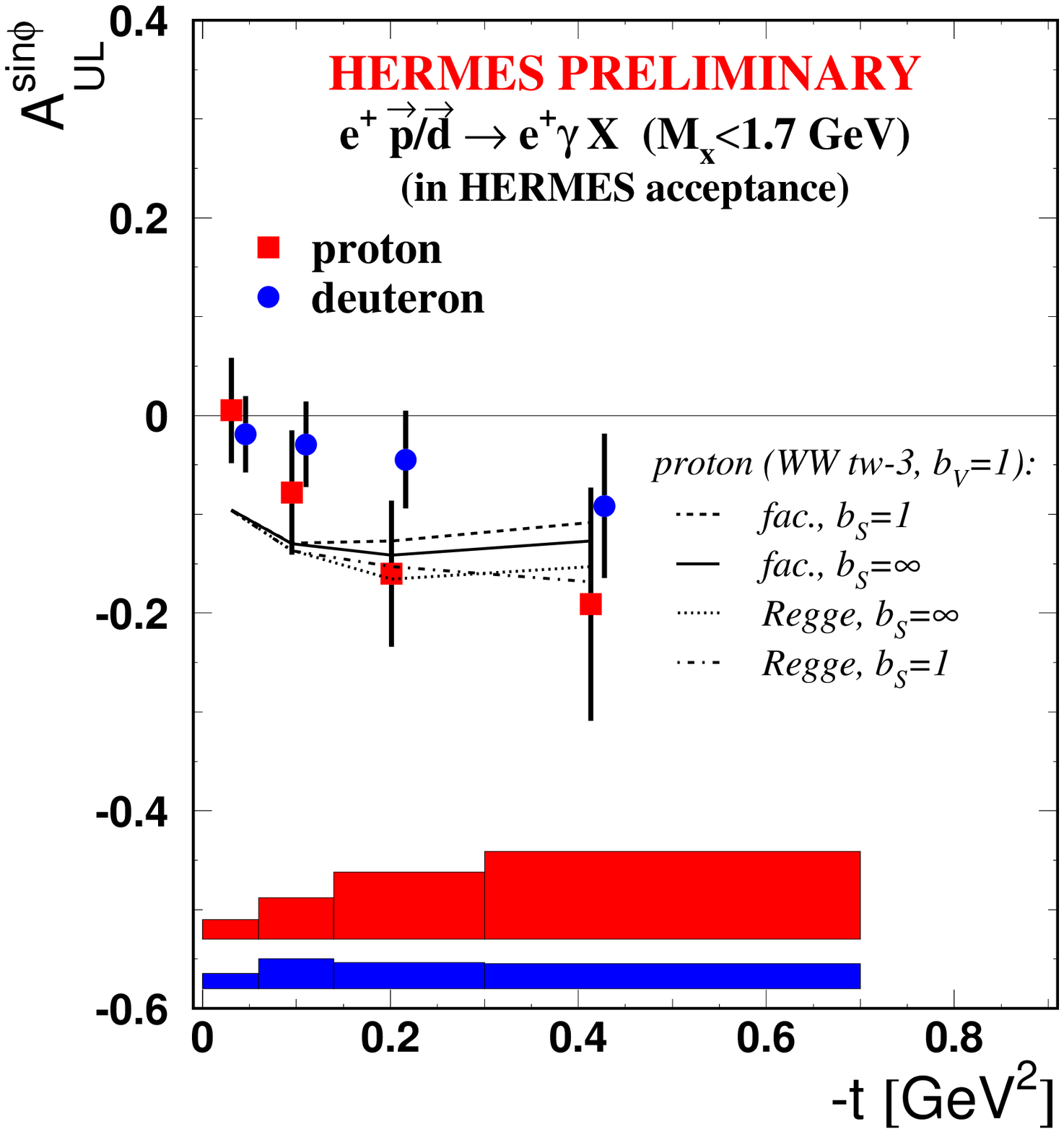}
  \includegraphics[height=.32\textheight]{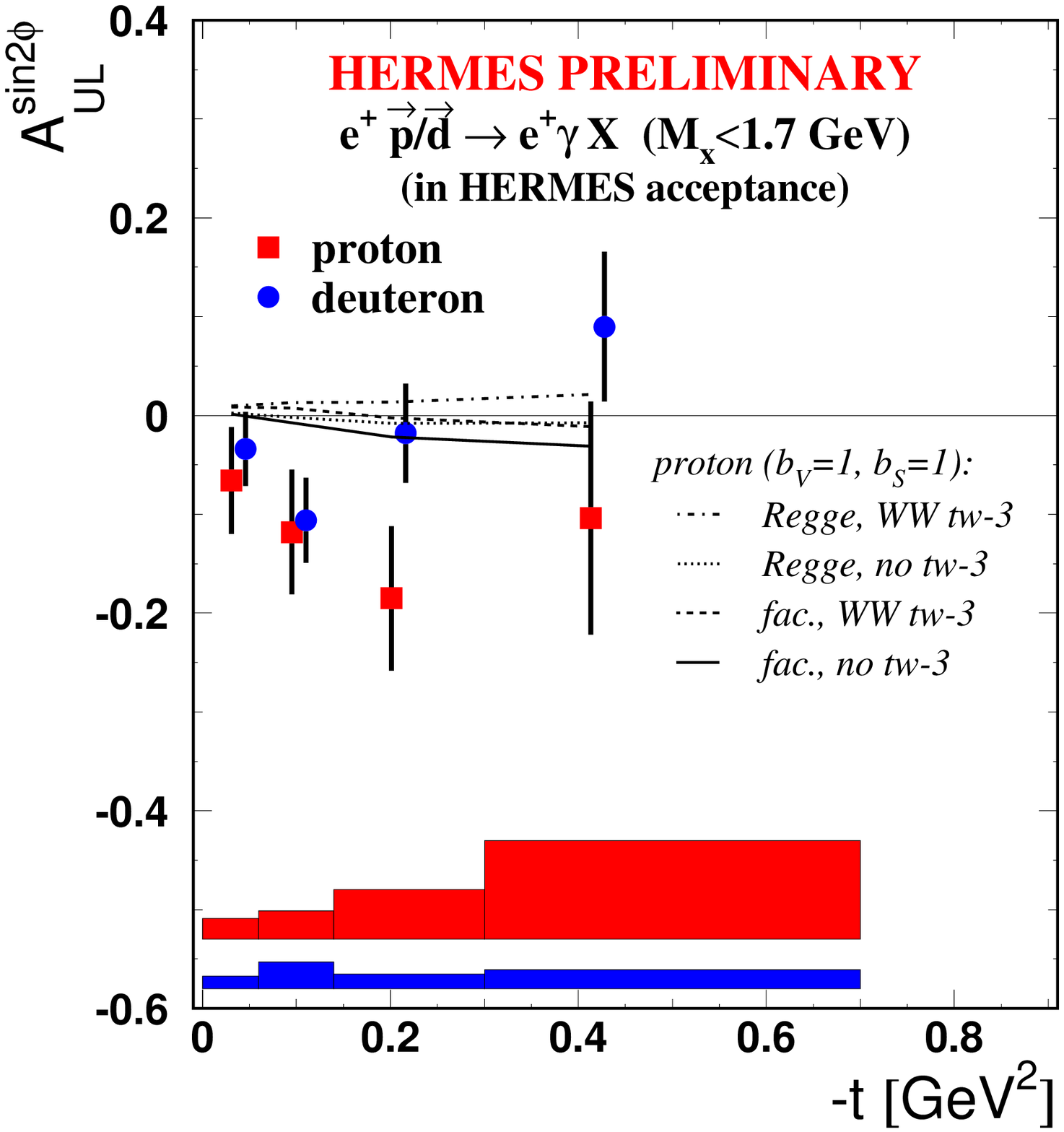}
  \caption{The $\sin \phi$ and $\sin 2 \phi$ moments 
           of longitudinal target-spin asymmetry 
           on hydrogen and deuterium as function of $t$
           as measured by HERMES  
           compared to predictions based on GPDs.}
  \label{fig:hdvcs}
\end{figure}

A review of the HERMES results on DVCS \cite{hermes:dvcs} has been presented, 
including new data on polarized  targets. On basis of unpolarized
target data, one can measure the beam charge asymmetries, which
are sensitive to the real part of the DVCS amplitudes, and the
beam spin asymmetries, which are sensitive to the imaginary part.
These are in fact mainly sensitive to the $H$ GPD. 
Both asymmetries have been extracted and show the expected 
$\cos(\phi)$ and $\sin(\phi)$ behavior, respectively. 
A measurement of the $t$ dependence of the beam charge asymmetry
has been performed and comparison with models indicate
the possible sensitivity of the data to constrain GPDs.
Polarized target have been analyzed and the longitudinal 
target spin asymmetry, which is sensitive to the $\tilde{H}$ GPD, has
been measured for the fist time. The resulting $\sin(\phi)$
and $\sin(2\phi)$ moments are shown as a function of $t$ in 
Fig.~\ref{fig:hdvcs}, together with prediction based on GPD models.
The sizeable $\sin(2\phi)$ moment might indicate a sensitivity
to the twist-3 $H$ and $\tilde{H}$ contributions. The installation
of a new recoil detector will allow to tag directly the final state 
proton and to reduce the uncertainties due to the 
backgrounds arising from  the missing mass techniques used up 
to now to guarantee exclusivity.

Gavalian \cite{jlab:dvcs} summarized the previous 
results on DVCS obtained by the CLAS
experiment at JLAB, which measured in particular the beam spin asymmetry.
In 2004 a dedicated DVCS experiment has been operated in Hall A and
new results are expected soon.
He also reviewed the status of the upgrade of the CLAS experiment 
which would allow to measure DVCS with the expected $12$ \gev\ beam.

Exclusive meson production processes provide as well access to GPDs.
HERMES has studied \cite{hermes:meson} exclusive \piplus\ production
which is sensitive to the $\tilde{H}$ and the $\tilde{E}$ GPDs. The
\qsq\ dependent cross section has been measured and found to be in
good agreement with a GPDs based model. A first measurements of the target
spin asymmetry for \rhoz\ production, which probes the $E$ GPD, has been
performed.

Weiss \cite{weiss} reviewed the theoretical status of hard electroproduction
of pions and kaons and their link to GPDs.

%%%%%%%%%%%%%%%%%%%%%%%%%%%%%%%%%%%%%%%%%%%%%%%%%%%%%%%%%%%%%%%%%%%%%%%%%%%%%%%

\section{Towards the LHC}

Diffractive physics has provided a rich source of important
results from both HERA and the TEVATRON.
Within the past few years there has been increasing interest to
the study of diffractive processes at the LHC in connection
with the  proposal to add forward proton detectors
to the LHC experiments.
Various aspects of physics with forward proton tagging
at the LHC have been under discussion in our working group. 

Eggert \cite{ke} described the status of the TOTEM detector and the prospects
of measurements of total and elastic $pp$ cross section. In particular,
the total $pp$ cross section will be measured with the record (order 1 \%)
accuracy. This would allow to strongly restrict the range of existing
theoretical models. Elastic cross section will be measured in the wide interval
of momentum transfer $10^{-3}~<~-t~<~8~$GeV$^2$.

The measurement of the elastic slope $b(t=0)$ at the LHC is especially important,
since it is expected 
(see for example \cite{KMRcetr,KMRsoft}) that this quantity
is much more sensitive to the effects of the multi-Pomeron cuts 
than the total cross section.

Studies of diffractive
physics at TOTEM require integration with CMS. CMS and TOTEM together will 
provide the largest acceptance detector ever built at a hadron collider.
From the point of view of testing different regimes of the asymptotical
behavior of the $pp$ scattering amplitude, it will be very informative to
measure accurately the survival probabilities of one, two, three (maybe even four)
rapidity gaps~\cite{Dino,goul,Prosp}.   
CMS/TOTEM physics menu will include also measurements of 
the centrally produced low mass systems ($\chi$-bosons, dijets, diphotons).
Special attention in his talk Eggert paid 
to the new ($\beta^*$  =172m ) optics
aimed at optimization of diffractive proton detection at 
L= 10$^{32}$cm$^{-2}$s$^{-1}$.

Several speakers  (Albrow, Cox, Kowalski, Piotrzkowski and Royon ) discussed the
unique physics potential of forward proton tagging at 420m at the LHC.
The use of forward proton detectors as a means to study Standard Model (SM) and new physics
at the LHC has only been fully appreciated within the last few years, (see 
e.g.~\cite{Prosp,cox1} and references therein).
By detecting protons that have lost less than 2 \%   of their longitudinal
momentum, a rich QCD, electroweak, Higgs and BSM program becomes accessible, with
a potential to study phenomena which are unique at the LHC, and difficult even at a future
linear collider~\cite{albr,cox,fp420,kow,kp}.

It was emphasized by Albrow, Cox and Royon \cite{cr} that
the so-called central exclusive production (CEP) process might provide a 
particularly clean environment to search for, and identify 
the nature of, new particles at the LHC.
There is also a potentially rich, more exotic physics menu 
including (light) gluino and squark production,
 gluinonia, radions, and indeed any object which has $0^{++}$ or $2^{++}$ quantum 
numbers and couples strongly to gluons \cite{Prosp}. 

%By central exclusive, we refer to the process 
% where the symbol meaning 
%has been described in the section ``Diffraction at the TEVATRON''.
%where the central system would consist of 2 $b$-quark jets, and {\it no other activity}. 
%The process is attractive for two main reasons. 
Central exclusive production processes $pp\rightarrow p \oplus \phi \oplus p$,
where 
$\oplus$ denotes the absence of hadronic activity ('gap') 
between the outgoing hadrons and the decay products of the central 
system $\phi$, are attractive for two main reasons. 
Firstly, if the outgoing protons remain intact and scatter
 through small angles, then, to a very good approximation, the central system 
$\phi$ must be dominantly produced in a spin $0$,
 CP even state, therefore allowing a clean determination of the quantum numbers
 of any observed resonance. Secondly, as a result of these quantum number selection 
rules, coupled with the (in principle) excellent mass resolution on the central
 system achievable if suitable proton detectors are installed, signal to
background ratios greater than unity are predicted for SM
Higgs production \cite{DeRoeck:2002hk}, and significantly larger for the 
lightest Higgs boson in certain regions of the MSSM parameter space 
\cite{Kaidalov:2003ys}. Simply stated, the reason for these large signal to 
background ratios is that exclusive $b$ quark production, the primary 
background in light Higgs searches, is heavily suppressed due to the quantum 
number selection rules. Another attractive feature is the ability to directly 
probe the CP structure of the  Higgs sector by measuring azimuthal asymmetries 
in the tagged protons 
\cite{Khoze:2004rc}. Another strategy to explore the manifestation of the
explicit CP violation in the Higgs sector was recently studied by Ellis et 
al.~\cite{je}.

\begin{figure}
  \includegraphics[height=.2\textheight]{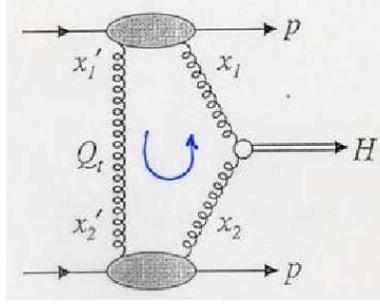}
  \caption{Schematic diagram for central exclusive Higgs production at the LHC, 
           $pp~\rightarrow~p~+~H~+~p$.}
  \label{fig:f17}
\end{figure}

The 'benchmark' CEP process for new physics
searches is SM Higgs production, sketched in Fig.~\ref{fig:f17}. The cross section prediction for the production of a 120 GeV  Higgs at 14 TeV is 
3 fb, falling to 1 fb at 200 GeV
\cite{Khoze:2000}. \footnote {for the
discussion of the uncertainties in this calculation, see \cite{Kaidalov:2003ys}.}
The simplest channel to observe the SM Higgs in the tagged proton approach 
from the experimental perspective is the $WW$ decay channel \cite{cox,cox2}.
More challenging from a trigger perspective in the  $b \bar b$ channel.
This mode, however, becomes extremely important in the so-called
'intense coupling regime' of the MSSM, where the CEP is
likely to be the discovery channel. In this case it is expected close to $10^{3}$
exclusively produced double-tagged Higgs bosons in 30fb$^{-1}$ of delivered luminosity.
About 100 would survive the experimental cuts, with a signal-to-background
ratio of order 10.

Furthermore, as was reported in \cite{sw,sk,kp}, 
forward proton tagging will make possible a unique program of
high energy photon interactions physics at the LHC. 
For example, the two photon production of $W$ pairs will allow a high precision study
of the quartic gauge couplings \cite{kp}.
Photon interactions are enhanced in heavy ion collisions and
studies of such ultra peripheral collisions were discussed in \cite{sw,sk}.
In addition, two photon exclusive production of lepton pairs
provides an excellent tool for calibrating both luminosity and the energy scale
of the tagged events \cite{kp}.

Finally, by tagging both outgoing protons, 
the LHC is effectively turned into a glue-glue collider.
This will open up a rich, high rate QCD physics menu (especially in what
concerns diffractive phenomena), allowing to study the skewed unintegrated
gluon densities and the details of rapidity gap survival \cite{kow}. 
Note that the CEP provides a source of practically
pure gluon jets (gluon factory \cite{factory}). This can be an ideal laboratory
to study the properties of gluon jets, especially in comparison with the
quark jets, and even the way to search for the glueballs.

Cox and Kowalski \cite{cox,kow} discussed the outline of the
FP420 R\&D project aimed at assessing whether it is 
possible to install forward proton detectors with appropriate acceptance 
at ATLAS and/or CMS, and to fully integrate such detectors within the 
experimental trigger frameworks \cite{fp420}.

\section{SUMMARY and OUTLOOK}
A wealth of diffractive data is available over an extended kinematic
range from the HERA experiments, allowing precise measurements of the
diffractive structure function and the extraction of the Diffractive
Parton Distribution Functions (DPDFs). The main news here is that, for the
first time, several independent next-to-leading-order QCD fits to the data
have become available. While all fits suggest that the DPDFs are
gluon-dominated, significant differences between the various sets are
evident. The origin of these differences is under investigation; 
at the moment they are the only information we have on the DPDFs 
uncertainties. 
A precise and consistent determination of the DPDFs is one of the main aims
of the HERA community in the near future. These functions provide an important
input for the prediction of inclusive hard diffractive cross sections at the
LHC. They are also essential to test the validity of the QCD
factorisation in diffractive processes.

Recent results on deeply virtual Compton scattering and exclusive
meson production from the HERA experiments and from COMPASS and CLAS are 
especially interesting because of their sensitivity to the Generalized 
Parton Distributions (GPDs). GPDs extend the standard PDFs by including the
information on correlations between partons and on their transverse
momentum.  GPDs would allow, through Ji's sum rule, the determination of
the contribution of the quark angular momentum to the proton spin.
Moreover, they provide a useful input for the prediction of the
diffractive exclusive cross sections at the LHC. In this respect, the
measurements of central exclusive production rates, on the way with run-II
TEVATRON data, play also an important role.

In the past few years there has been an increasing interest in the study
of diffractive processes at the LHC. The TOTEM detector, integrated with
CMS, will allow the study of diffractive physics with the largest
acceptance detector ever built at a hadron collider. 
The LHC physics programme can be significantly enlarged by
equipping a region at 420~m from the ATLAS and/or
CMS interaction point, as has been recently proposed by the FP420 R\&D 
project. This would provide 
a particularly clean environment to search for, and to identify the nature 
of, the new particles (first of all, the Higgs boson). At the same time
this would open a rich QCD menu and allow a unique program
of high-energy photon interactions.

\begin{theacknowledgments}

We would like to thank all the people involved in the conference preparation 
for the perfect organization and the warm atmosphere and all the speakers 
and participants of our working group for their contribution to talks and discussion. 
We are especially grateful to Wesley Smith for his continuous support before, 
during and after the conference. 

\end{theacknowledgments}

\end{document}